\documentclass[useAMS,usenatbib]{mn2e}
\usepackage{epsf}
\usepackage{amssymb}
\usepackage{graphicx}
\usepackage{eqnarray,amsmath}
\usepackage{bigfoot}
\usepackage[usenames]{color}
\usepackage{placeins}

\newcommand {\bc}{\begin {center}}
\newcommand {\ec}{\end {center}}
\newcommand {\be}{\begin {equation}}
\newcommand {\ee}{\end {equation}}
\newcommand {\Chandra}{{\it Chandra }}
\newcommand {\XMM}{{\it XMM-Newton }}
\newcommand {\XIPE}{{\it XIPE}}

\def\deg{$^{\circ}$}

\title[Sgr A* X-ray echo]{Polarization and long-term variability of Sgr A* X-ray echo}
\author[Churazov et al.]{E.~Churazov$^{1,2}$,  I.Khabibullin$^{1,2}$, G.~Ponti$^{3}$, R.~Sunyaev$^{1,2}$\\
$^{1}$ MPI f\"ur Astrophysik, Karl-Schwarzschild str. 1, Garching D-85741, Germany\\
$^{2}$ Space Research Institute, Profsoyuznaya str. 84/32, Moscow
  117997, Russia\\
$^3$ MPI f\"ur extraterrestrische Physik, Giessenbachstrasse 1, Garching D-85748, Germany 
}

\begin{document}

\date{Accepted 2017 February 16. Received 2017 January 10; in original form 2016 December 01}

\pagerange{\pageref{firstpage}--\pageref{lastpage}} \pubyear{2012}

\maketitle

\label{firstpage}

\begin{abstract}
  We use a model  of the
  molecular gas distribution within $\sim 100$~pc from the centre
  of the Milky Way (Kruijssen, Dale \& Longmore) to simulate time evolution and polarization
  properties of the reflected X-ray emission, associated with the past
  outbursts from Sgr~A$^*$. While this model is too simple to describe
  the complexity of the true gas distribution, it illustrates the
  importance and power of long-term observations of the reflected
  emission. We show that the variable part of X-ray emission observed by \Chandra
  and \XMM from prominent molecular clouds is well described by a pure
  reflection model, providing strong support of the reflection
  scenario. While the identification of Sgr~A$^*$ as a primary source for this reflected emission is
  already a very appealing hypothesis, a decisive test of this model can be provided by future X-ray polarimetric observations, that will allow placing constraints on the location of the primary source.
  In addition, X-ray polarimeters
 (like, e.g., \XIPE) have sufficient sensitivity to constrain the
  line-of-sight positions of molecular complexes, removing major
  uncertainty in the model.
\end{abstract}

\section{Introduction}
\label{sec:intro}
Several molecular complexes in the central $\sim 100$~pc of the Milky
Way are not only prominent IR or submm sources, but they are also
bright in X-rays \citep[see, e.g.,][for review]{2010RvMP...82.3121G}. Their spectra have several features characteristic
for X-ray reflection by cold gas. It was suggested that Sgr~A$^*$ (a
supermassive black hole at the centre of the Galaxy) is responsible
for the illumination of these clouds
\citep[e.g.,][]{1993ApJ...407..606S,1993Natur.364...40M,1996PASJ...48..249K}. This
suggestion implies that Sgr~A$^*$, despite its present-day low
luminosity \citep[see, e.g.,][for review]{2014ARA&A..52..529Y},
was much brighter few hundred year ago. Since then, the problem of
clouds illumination has been subject to many theoretical and
observational studies \citep[see][for
  review]{2013ASSP...34..331P}. One of the specific predictions of the
illumination/reflection scenario is that the continuum emission is
polarized \citep{2002MNRAS.330..817C}. The polarization can not
only constrain the position of a primary source illuminating a given
cloud (from the orientation of the polarization plane), but also the
position of the cloud along the line of sight (from the degree of
polarization).

There were many studies that consider the reflection scenarios in the
GC region \citep[e.g.,][]{1998MNRAS.297.1279S,2000ApJ...534..283M,2014sf2a.conf...85C,2016A&A...589A..88M,2015A&A...576A..19M}. Here we return to this problem to
emphasize the power of long-term monitoring of the reflected
emission. We illustrate this point by using models of the molecular
gas distribution in the central $\sim$100 pc around Sgr A$^*$
\citep[e.g.,][]{Molinari2011,Kruijssen2015,Henshaw2016,2016A&A...586A..50G,Schmiedeke2016}
to calculate the expected reflected signal over few hundred years. Even
although the true distribution of the gas is likely more complicated
than prescribed by these models \citep[see][]{churazov16}, this
exercise shows that mapping molecular gas by X-ray observations will
eventually lead to a clear 3D picture of the central molecular
zone. We show that new X-ray polarimetry missions that are currently
under study by ESA and NASA
\citep[e.g.,][]{2013ExA....36..523S,2013SPIE.8859E..08W} have sufficient
sensitivity  to make a decisive contribution in
resolving the ambiguity in the position of molecular clouds along the
line of sight, and consequently confirm or reject the identification of Sgr A$^*$ as a primary source of the illumination \citep[see also ][]{2015A&A...576A..19M}.

Throughout the paper we assume the distance to Sgr~A$^*$
 $R_0=8~{\rm kpc}$ \citep[see, e.g.,][for discussion of different distance indicators]{2010RvMP...82.3121G}, therefore $1'$ corresponds to 2.37~pc.

\section{Spherical cloud illuminated by a steady source}
\label{sec:modela}
We first consider a simple case of a uniform spherical cloud
illuminated by a steady source and we perform full Monte Carlo simulations of
the emergent spectrum, taking into account multiple scatterings,
similar to \citet{2008MNRAS.385..719C}.

\subsection{Illuminating source spectrum}
We assume that the primary unpolarized X-ray source is located at a
large distance from the cloud (much larger than the size of the cloud)
and, therefore, the cloud is illuminated by a parallel beam of
radiation, i.e. incoming photons are hitting a hemisphere facing the
primary source (see Fig.~\ref{fig:geo} in Appendix \ref{ap:model}). The illuminating source has a power law spectrum with
a photon index $\Gamma=2$, i.e., $I(E)=A E^{-\Gamma}~{\rm
  photons~s^{-1}~keV^{-1}}$, where $E$ is the photon
energy. Since we are interested primarily in the reflected spectrum in
the 1-10 keV band, the behavior of the spectrum at high energies is
not very important for this study.

\subsection{Radiative transfer}
\label{sec:rad}
In modelling the radiative transfer, we followed the assumptions used
in \citet{2008MNRAS.385..719C}. In particular, we have included the
following processes: photoelectric absorption, fluorescence, Rayleigh
and Compton scattering. The abundance of heavy elements in the
molecular gas is highly uncertain
\citep[e.g.,][]{2007PASJ...59S.221K}. To model photoelectric
absorption we used solar abundance of heavy elements from
\citet{1992PhyS...46..202F}, but set the abundance of iron to 1.6
solar \citep[see, e.g.,][]{2010PASJ...62..423N,2010ApJ...719..143T,2015ApJ...814...94M}. The cross sections
for photoelectric absorption were taken from approximations of
\citet{1995A&AS..109..125V} and \citet{1996ApJ...465..487V}. For calculation of
the fluorescent emission we used fluorescent yields from
\citet{1993A&AS...97..443K}. The fluorescent lines of all
astrophysically abundant elements were included in the calculations
and added to the corresponding bins of the scattered spectrum. The
fluorescent photons are emitted isotropically without any net
polarized. The modeling of the Rayleigh and Compton scattering
employs differential cross-sections from the GLECS package
\citep{2004NewAR..48..221K} of the GEANT code
\citep{2003NIMPA.506..250A} and the Klein-Nishina formula for free
electrons. This model effectively accounts for effects of bound
electron on the shape of the Compton shoulder of scattered fluorescent
lines \citep[see ][]{1996AstL...22..648S,1998AstL...24..271V},
leading to a less sharp peak of the backscattered emission.

The photons coming from the direction of the primary source are
absorbed or scattered in the cloud. A weight $w=1$ is assigned to each
incoming photon. The evolution of each photon is traced for multiple
scatterings inside the cloud, each time reducing the weight according
to the probability of being absorbed or escape from the cloud after
each scattering. The process stops once the weight drops below a
certain threshold, in our case $w<10^{-8}$.

\subsection{Model spectra}
\label{sec:rmod}
With the above assumption the emergent spectrum depends on two
parameters: i) the optical depth of the cloud, parametrized via the
Thomson optical depth of the cloud $\tau_T=\sigma_T n_H R$, where
$n_H$ is the number density of hydrogen atoms and $R$ is the radius of
the cloud, and ii) $\mu=\cos \theta$, the cosine of the angle $\theta$
between the direction of the primary beam and the line of sight (see Fig.~\ref{fig:geo} in Appendix \ref{ap:model}). In
the code the total flux of the of incoming photons is normalised to
unity (in ${\rm photons~s^{-1}~keV^{-1}}$) at 1 keV. 

\begin{figure*}
\begin{minipage}{0.49\textwidth}
\includegraphics[trim= 1mm 6cm 0mm 2cm, width=1\textwidth,clip=t,angle=0.,scale=0.9]{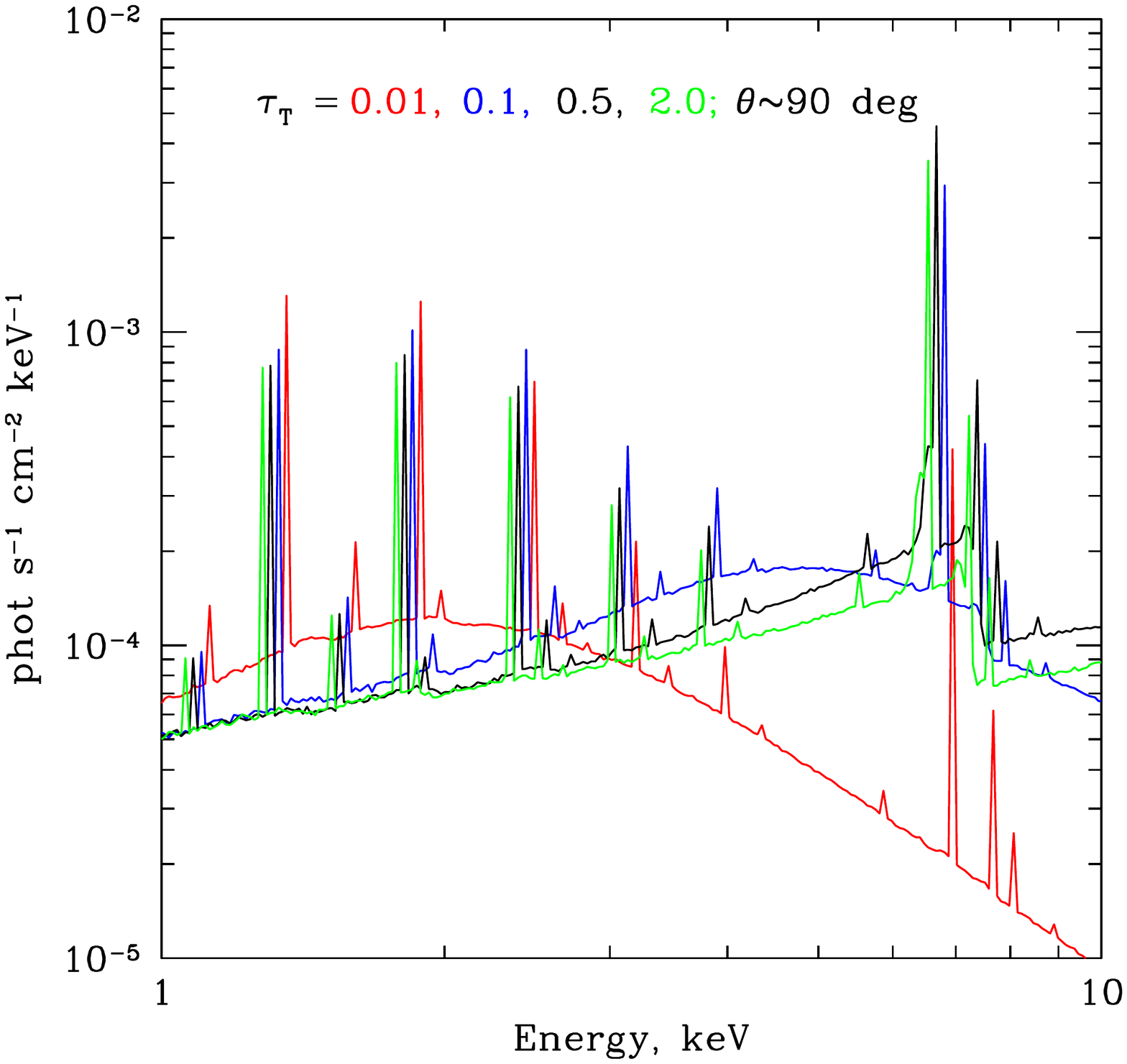}
\end{minipage}
\begin{minipage}{0.49\textwidth}
\includegraphics[trim= 1cm 6cm 0mm 2cm,width=1\textwidth,clip=t,angle=0.,scale=0.9]{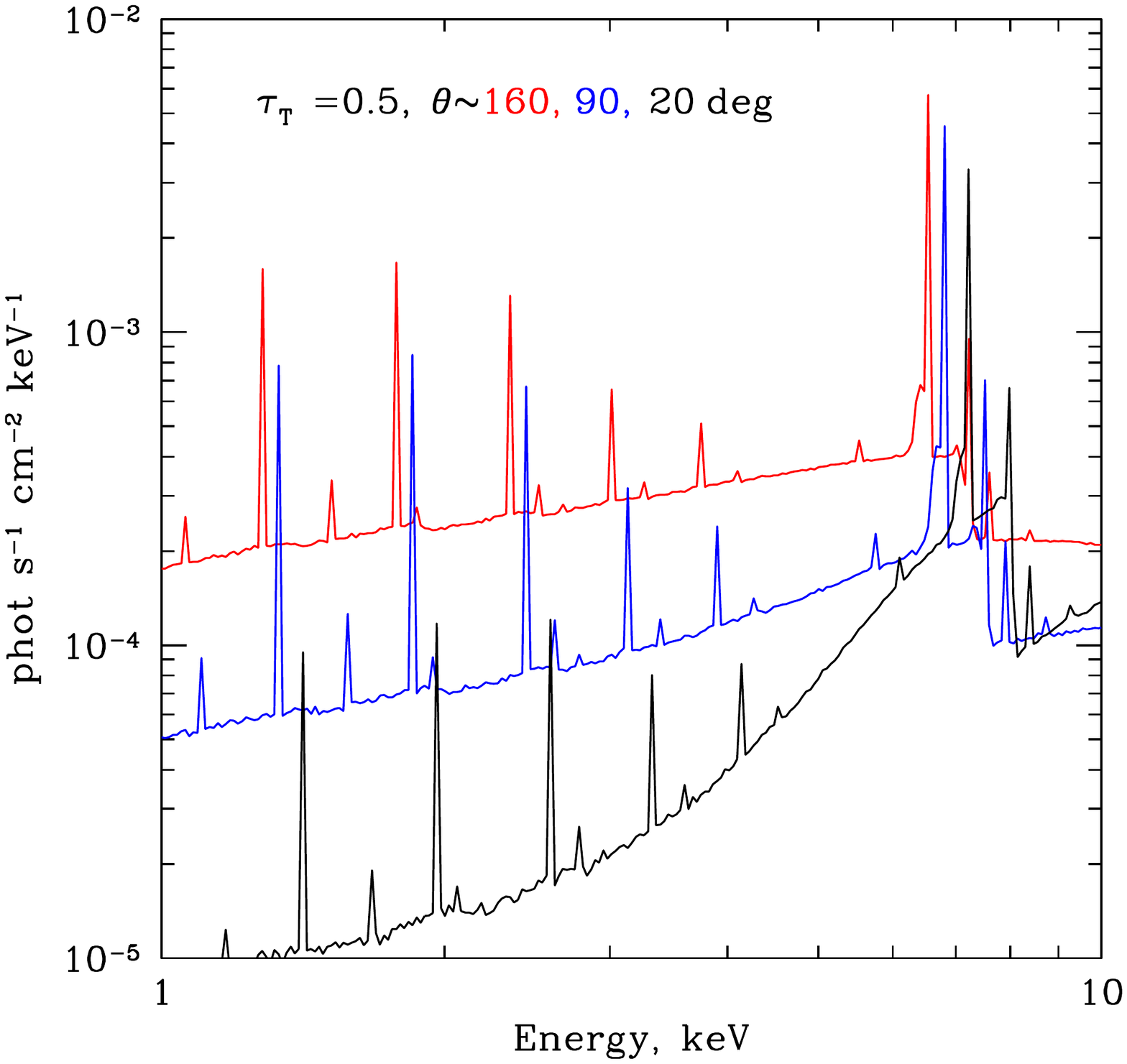}
\end{minipage}
\caption{Typical spectra emerging from a spherical could illuminated
  by a distant source with a power law X-ray spectrum (see
  \S\ref{sec:modela}).  Each plotted spectrum is shifted in energy for
  clarity. The most prominent fluorescent line is due to iron at
  $\sim$6.4~keV, while at lower energies the lines of Ca [3.7 keV], Ar
  [3.0 keV], S [2.3 keV], Si [1.7 keV] and Mg [1.2 keV] are visible,
  as well as few weaker fluorescent lines of less abundant elements.
  A ``Compton shoulder'' due to the scattered fluorescent photons is
  strong enough to be visible only for iron.  {\bf Left:} Dependence
  of the emergent spectrum on the optical depth of the cloud. The red,
  blue, black and green lines correspond to $\tau_T=0.01$, 0.1, 0.5
  and 2, respectively.  The angle between the line of sight and the
  primary beam is 90\deg\! for all spectra. For small optical depth
  ($\displaystyle \tau_T=0.01$) the spectrum (the red line) has a
  shape resembling the shape of the incident spectrum (except at low
  energies, where photoelectric absorption strongly dominates Thomson
  scattering). As the optical depth increases the spectrum evolves
  towards typical ``reflection'' spectrum from a semi-infinite
  medium. {\bf Right:} Dependence of the emergent spectrum on the
  angle $\theta$ between the line of sight and the primary beam. The
  optical depth ($\tau_T=0.5$) is the same for all spectra. For
  $\theta\sim$160\deg (red line), i.e. back scattering, the spectrum resembles
  reflection from a semi-infinite medium. In the opposite limit
  $\theta\sim$20\deg, i.e., a cloud illuminated from the back, the
  spectrum has a strong decline at low energies, caused by the
  photoelectric absorption (black line).
\label{fig:spec_ma}
}
\end{figure*}

Typical simulated emergent spectra (as functions of $\tau_T$ and
$\theta$) are shown in Fig.~\ref{fig:spec_ma}. For clarity each plotted
spectrum is slightly shifted in energy. The left panel shows the
variations of the observed spectrum with the optical depth of the
cloud, when the angle between the line of sight and the primary beam is
fixed at 90\deg. For small optical depth ($\displaystyle \tau_T=0.01$) the
spectrum (the red line) has a shape resembling the shape of the
incident spectrum (except at low energies, where photoelectric
absorption strongly dominates Thomson scattering). As the optical
depth increases the spectrum evolves towards a typical ``reflection''
spectrum from a semi-infinite medium. The right panel shows the
variations of the spectrum with the viewing angle, while the optical
depth is fixed to $\tau_T=0.5$. In this case, the scattering by
$\sim$160\deg, i.e. back scattering, leads again to the spectrum
resembling reflection from semi-infinite medium with a flat (and even
slightly raising with energy) spectrum (the red line). In the
small-angle scattering limit, i.e., we view a cloud illuminated from
the back, the spectrum has a stronger decline at low energies, caused
by the photoelectric absorption (the black line).

In section \S\ref{sec:comp} we use these spectra in application to the real
data to single out the contribution of the reflected emission to the
observed X-ray spectra in the GC region.

An {\small XSPEC}-ready model {\small ``CREFL16''} based on
\S\ref{sec:rad} is publicly available (see Appendix \ref{ap:model} for
a short description). The model covers 0.3-100 keV energy range. Apart
from the normalization, the model has four parameters: radial Thomson
optical depth of the cloud, slope of the primary power law spectrum,
abundance of heavy elements and the cosine of the viewing
angle. Compared to recently released model of
\citet{2016MNRAS.463.2893W}, the model includes fluorescent lines not
only for iron, but for other elements too and has abundance of heavy
elements (heavier than He) are a free parameter.

\section{Hints from observations}
\label{sec:comp}
Shown in Fig.~\ref{fig:xmm_maps} is the \XMM {\it EPIC-MOS}\footnote{European Photon Imaging Camera - Metal Oxide Semi-conductor CCD array} $\sim
110'\times 30'$ image of the GC region in the 4-8 keV band. The image
was accumulated over all publicly available observations in the
archive.  In this image, Sgr A$^*$ is located in a red patch
near to the centre of the image. In this section we will use the data from this
$\sim 110'\times 30'$ ($\sim 260\times 70$~pc) area for spectral
analysis of the diffuse emission and decomposition of the spectrum
into two components \citep[see also ][for the corresponding emission in X-ray continuum and soft X-ray lines]{2015MNRAS.453..172P}.

\subsection{Contributions of the reflected and hot plasma components}
\label{sec:2t}
\begin{figure*}
\begin{minipage}{0.9\textwidth}
\includegraphics[trim= 0mm 0cm 4.5cm 0cm, width=1\textwidth,clip=t,angle=0.,scale=0.9]{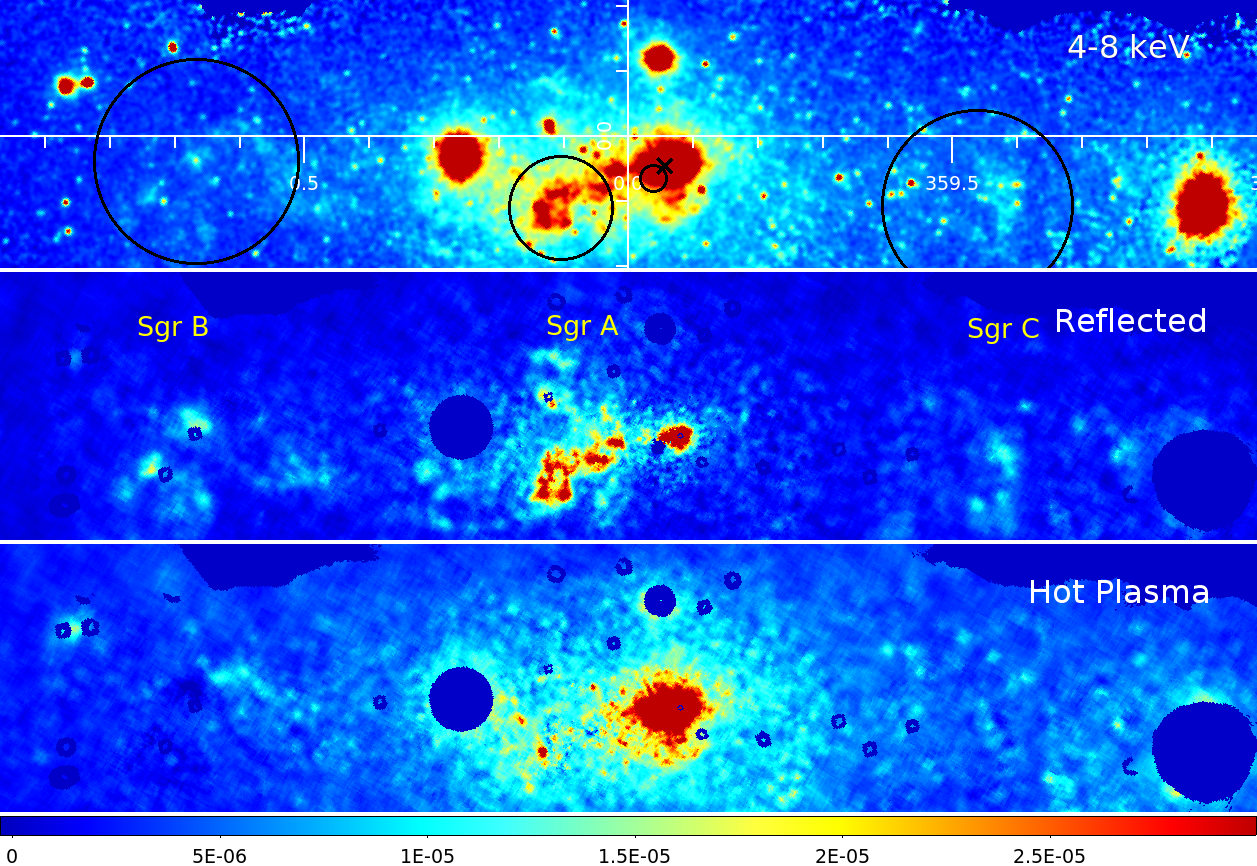}
\end{minipage}
\caption{\XMM maps ($\sim 110'\times 30'$) of the GC region in
  Galactic coordinates. The top panel shows the 4-8 keV map in units of
  ${\rm phot~s^{-1}~cm^{-2}~arcmin^{-2}}$; see colorbar at the bottom
  of the Figure. The position of Sgr~A$^*$ is marked with a black {\bf
    X} in the top panel. Several very bright spots correspond to
  bright compact sources (X-ray binaries), which appear very extended
  in the image that is tuned to show faint diffuse emission. Note also, that scattering by dust grains could lead to extended halos around sources in the Galactic Center region \citep[see, e.g.,][for the analysis of AXJ1745.6-2901 halo]{jin16}.
  The four black circles show the regions used for
  spectra extraction and analysis in \S\ref{sec:2t}. The three bigger
  circles correspond to the regions where the reflection component is
  known to be strong \citep[e.g.,][]{2014IAUS..303...94S}. The
  smallest circle represents an example of an X-ray bright region with
  a weak reflected component. The lower two panels show the
  decomposition (see \S\ref{sec:2m}) of the observed flux into
  reflected component (middle panel) and hot plasma component (bottom
  panel). In these panels the regions contaminated by bright X-ray
  binaries have been excluded - see ``holes'' in the images. The units
  are the same as for the top panel.  From the comparison of the two
  lower panels it is clear that the reflected component is much more
  irregular than the hot plasma component, manifesting the
  inhomogeneity of the molecular gas distribution.
\label{fig:xmm_maps}
}
\end{figure*}

\begin{figure*}
\begin{minipage}{0.24\textwidth}
\includegraphics[trim= 1cm 4cm 0mm 2cm,width=1\textwidth,clip=t,angle=0.,scale=0.9]{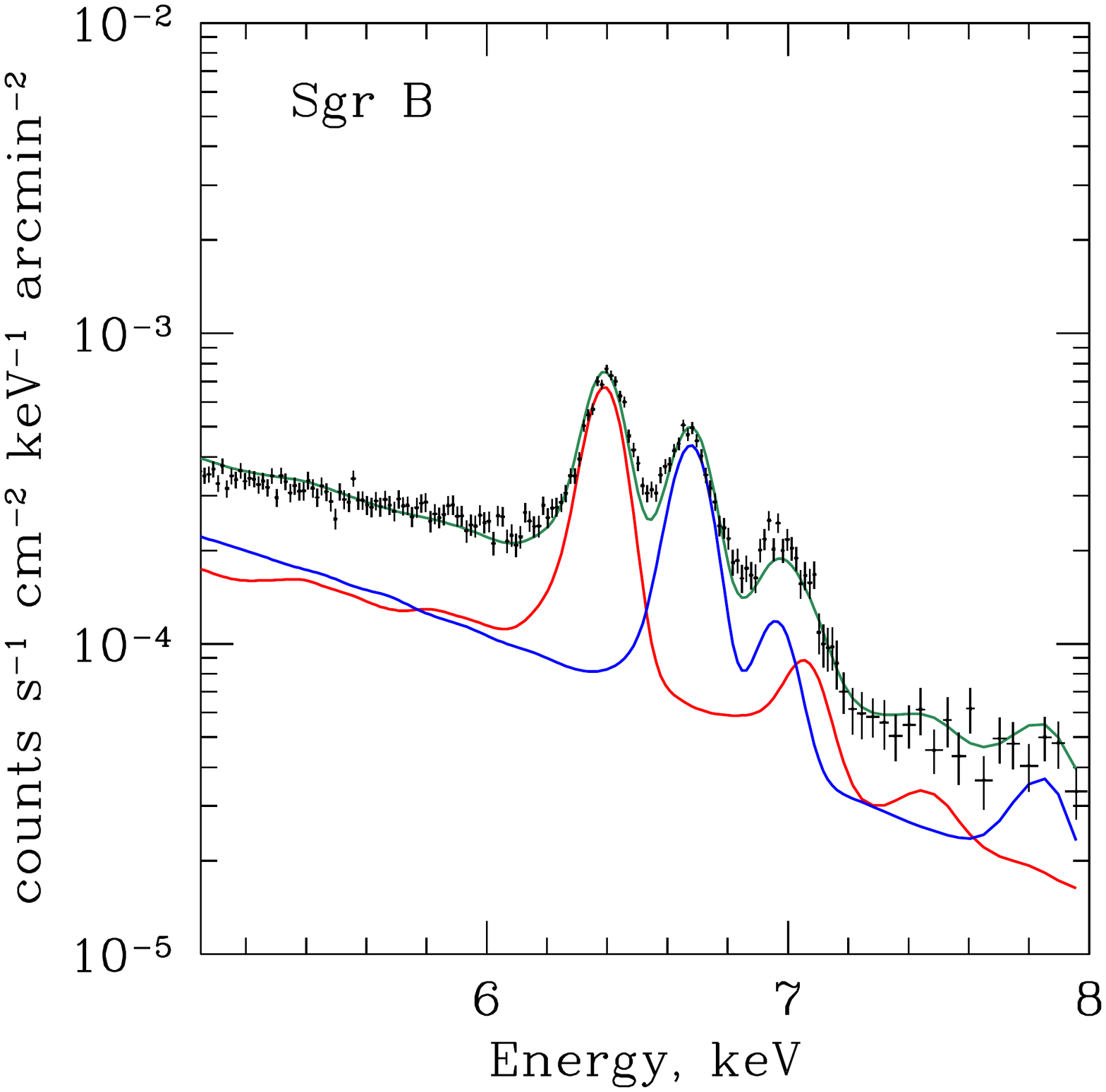}
\end{minipage}
\begin{minipage}{0.24\textwidth}
\includegraphics[trim= 1cm 4cm 0mm 2cm,width=1\textwidth,clip=t,angle=0.,scale=0.9]{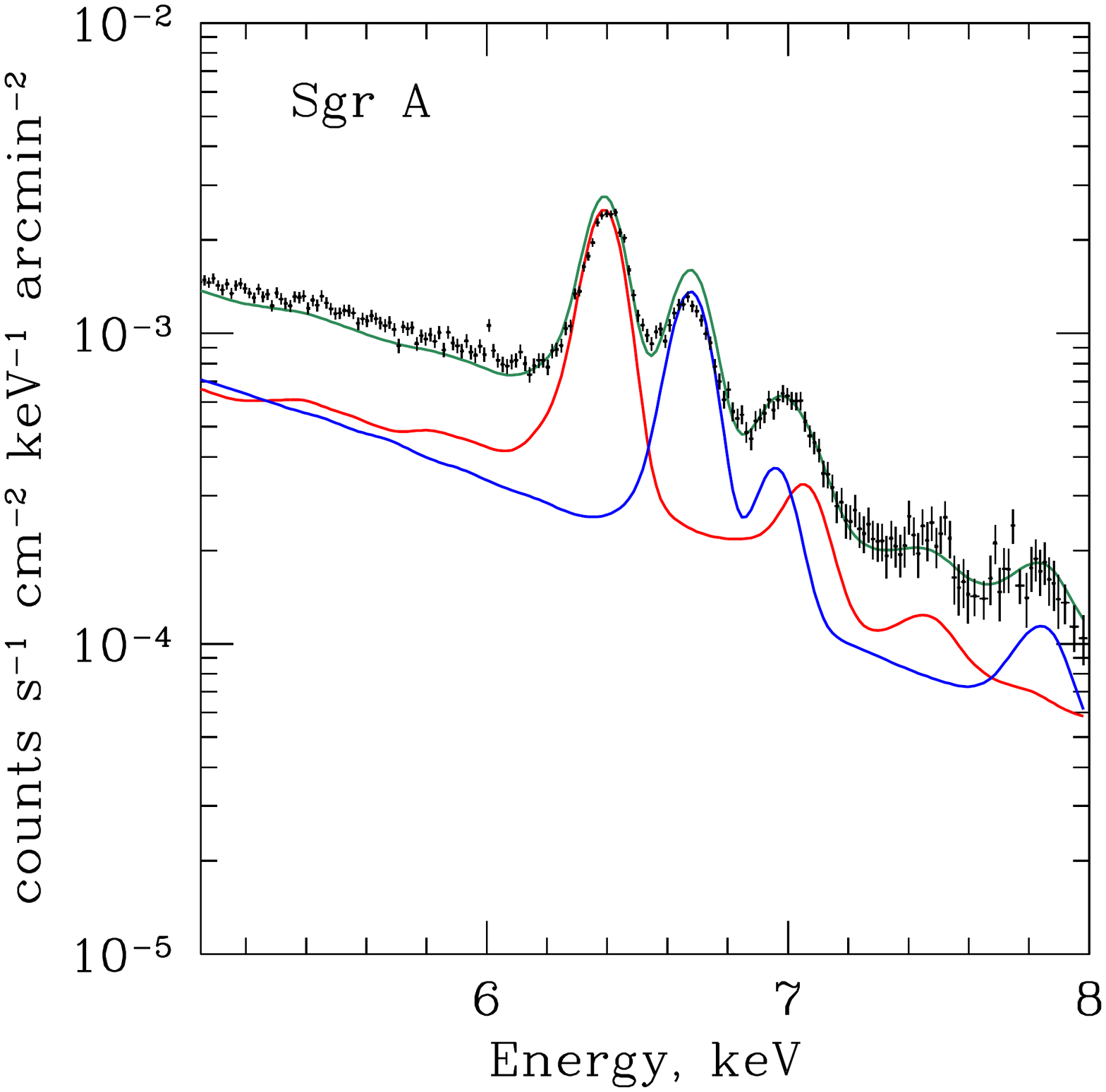}
\end{minipage}
\begin{minipage}{0.24\textwidth}
\includegraphics[trim= 1cm 4cm 0mm 2cm,width=1\textwidth,clip=t,angle=0.,scale=0.9]{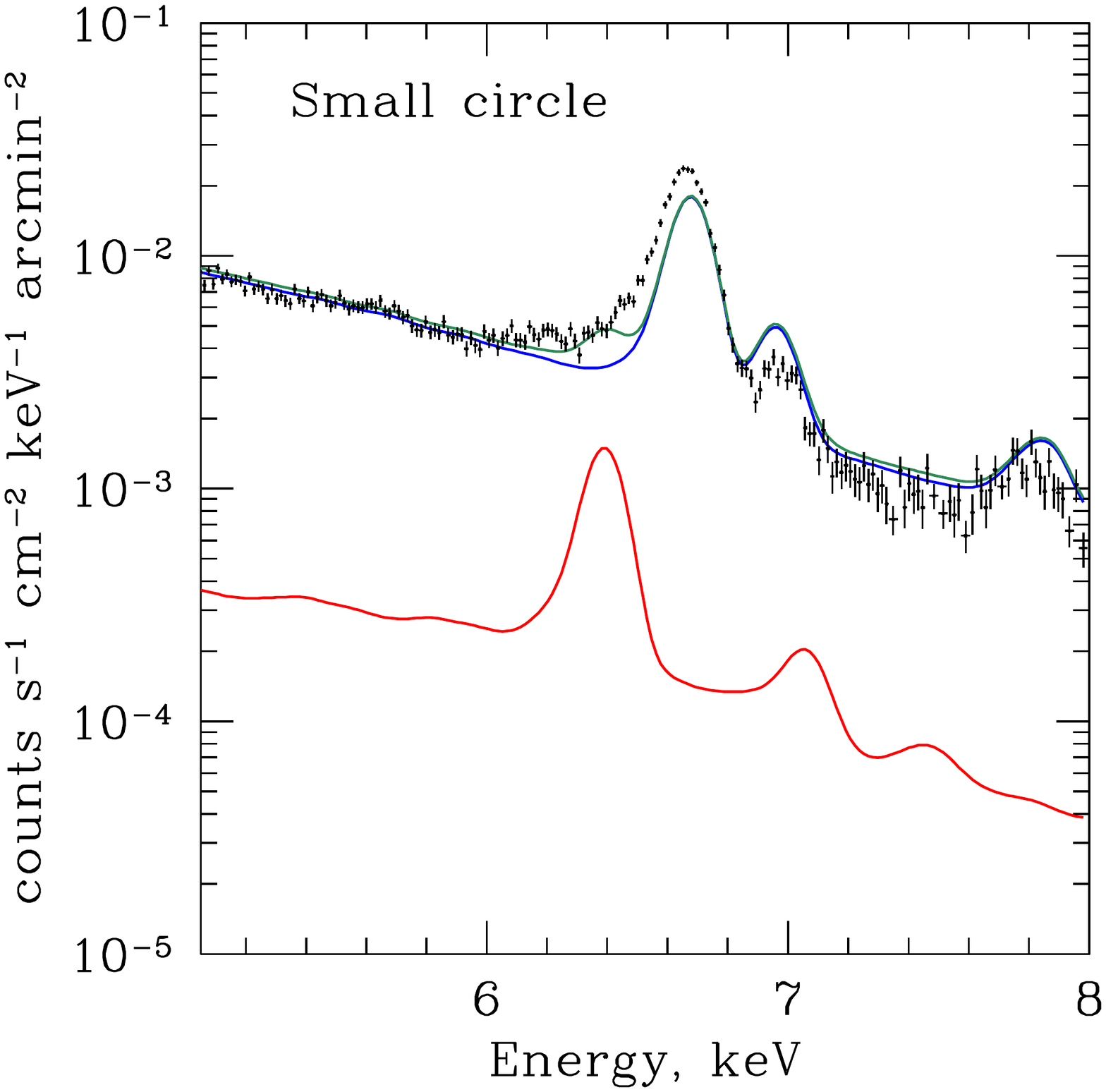}
\end{minipage}
\begin{minipage}{0.24\textwidth}
\includegraphics[trim= 1cm 4cm 0mm 2cm,width=1\textwidth,clip=t,angle=0.,scale=0.9]{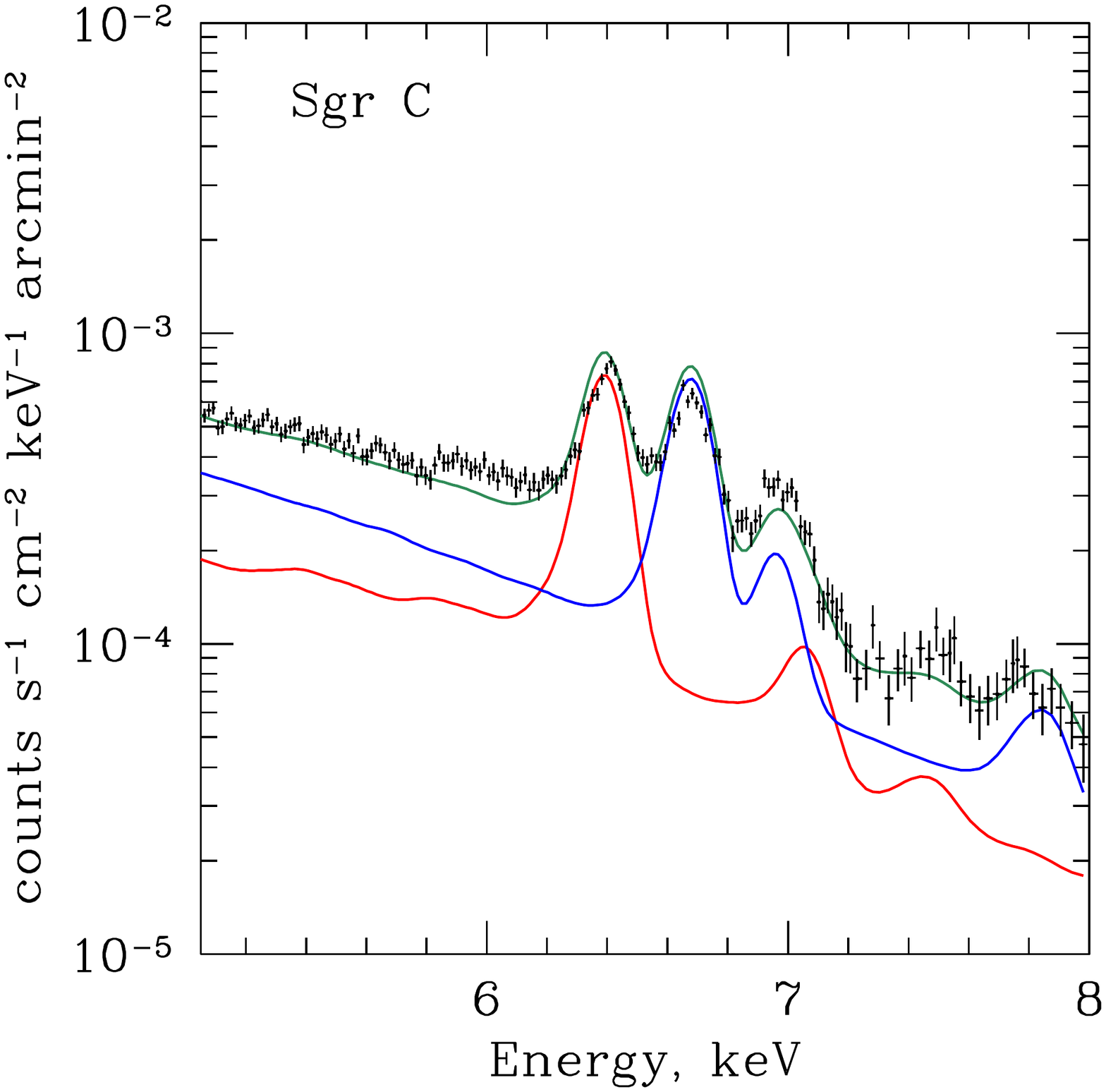}
\end{minipage}
\caption{The spectra (black crosses) in the 5-8 keV band extracted
  from the four circular regions shown in Fig.~\ref{fig:xmm_maps}. These
  spectra are modelled as a linear combination of two spectral
  templates: $kT=$6~keV optically thin plasma emission (blue) and
  reflected emission for $\tau_T=0.5$ and $\theta=90$\deg (red
    line, see \S\ref{sec:rmod}). Only the normalizations of both
  components are free parameters of the model. The green line shows
  the sum of the two components. In the three regions, where the reflected
  component is strong, the contributions of both components to the
  continuum are comparable.
\label{fig:xmm_4reg}
}
\end{figure*}

For spectrum extraction we have selected four regions shown with black
circles in Fig.~\ref{fig:xmm_maps}. Three bigger circles cover regions
known to be bright in the reflected radiation, from left to right: Sgr
B2, the ``Bridge'' and Sgr C \citep[see,
  e.g.,][]{2014IAUS..303...94S}. The resulting spectra, after excising
contributions from bright point sources, are shown in
Fig.~\ref{fig:xmm_4reg} with black crosses. The fourth (smaller) circle
is chosen, rather arbitrarily, to represent an example of X-ray bright
region with a weak reflected component. All four spectra show two very
prominent X-ray lines at 6.7 and 6.9 keV that have been identified
with the lines of He-like and H-like ions of iron, produced by hot
plasma \citep[e.g.,][]{1989Natur.339..603K}. Despite of the diffuse
appearance of this component, it can, at least partly, be attributed to 
the emission coming from compact sources - accreting
white dwarfs as in the Galactic Ridge \citep[see][]{2009Natur.458.1142R}, while the fractional contribution of compact source may vary across the region considered here. The three spectra the from
regions known to be bright in reflected emission, show, in addition, a
very prominent line at $\sim$6.4 keV, corresponding to fluorescent
emission of neutral (or weakly ionized) iron.

Assuming that the diffuse X-ray emission in the GC regions can be 
produced by the sum of hot plasma emission and reflected emission, we try to
model the observed spectra as a linear combination of these two spectral
templates. We reproduce the hot plasma component with the APEC model
\citep{2001ApJ...556L..91S} of an optically thin plasma emission with
a temperature fixed to 6~keV \citep[e.g.][]{2007PASJ...59S.237K}. For the reflected component we have
selected one of the simulated spectra of the reflected emission,
namely the one corresponding to $\tau_T=0.5$ and the scattering angle
of 90\deg. While these two templates are not able to describe the full
complexity of the observed spectra, the hope is that over a limited
energy range from $\sim$5 to $\sim$8 keV these two templates might
capture essential signatures of both components. We therefore fit the
observed spectra with the sum of these templates and we determine their
normalizations, while keeping all other parameters fixed. We use:
\begin{eqnarray}
  S(E)=A_1 R(E)+A_2 P(E),
\end{eqnarray}
where $S(E)$ is the observed spectrum, $R(E)$ and $P(E)$ are the
template spectra for the reflected component and for hot plasma emission,
respectively, and $A_1$ and $A_2$ are the free parameters of the
fit. The resulting fits are shown in Fig.~\ref{fig:xmm_4reg}. The red
and blue lines show the reflected and hot plasma components and the
green line shows the sum. Despite the simplicity of the model, the
best fit can reproduce the essential features of the spectrum. For each of the three regions with a
prominent reflection component the fit clearly shows that the reflected and hot plasma components contribute roughly equally to the continuum.
Later, in \S\ref{sec:xipe}, we will use this result in
order to estimate the polarization of the total emission. Fig.~\ref{fig:xmm_4reg} shows that for the forth
region the quality of the fit is worse and the contribution of the
reflected component is anyway small. We used this region only for the
sake of a test, to compare with the results for three other regions.

\subsection{Maps of the reflected emission}
\label{sec:2m}
Given the success of the template fitting in selected regions (see,
\S\ref{sec:2t}), we use the same approach to make a map of the
reflected component. Since our model is linear, we can easily do the
decomposition by calculating in every pixel of the image the scalar
products of the observed spectra with the two spectral templates in
the 5-8 keV band, and solve for the best-fitting amplitudes. When
calculating scalar products, the template spectra have been convolved
with the instrument spectral response. The maps of the two components
(best-fitting amplitudes for each template) derived from
\XMM data are shown in the two lower panels of
Fig.~\ref{fig:xmm_maps}.  The maps have been adaptively smoothed to
reach the required level of statistics (number of counts in the 5-8~keV
band per smoothing window).

The resulting maps appear qualitatively similar to the 6.4 keV maps
derived in \citep[e.g.][]{2014IAUS..303...94S,2014IAUS..303..333P}, confirming that this simple approach
can be used to reveal the morphology of the reflected emission and
estimate its flux. A more elaborate spectral analysis is done below.

\subsection{Difference spectra}
\label{sec:dif}
\begin{figure*}
\begin{minipage}{0.49\textwidth}
\includegraphics[trim= 0mm 0cm 6cm 0cm, width=1\textwidth,clip=t,angle=0.,scale=0.65]{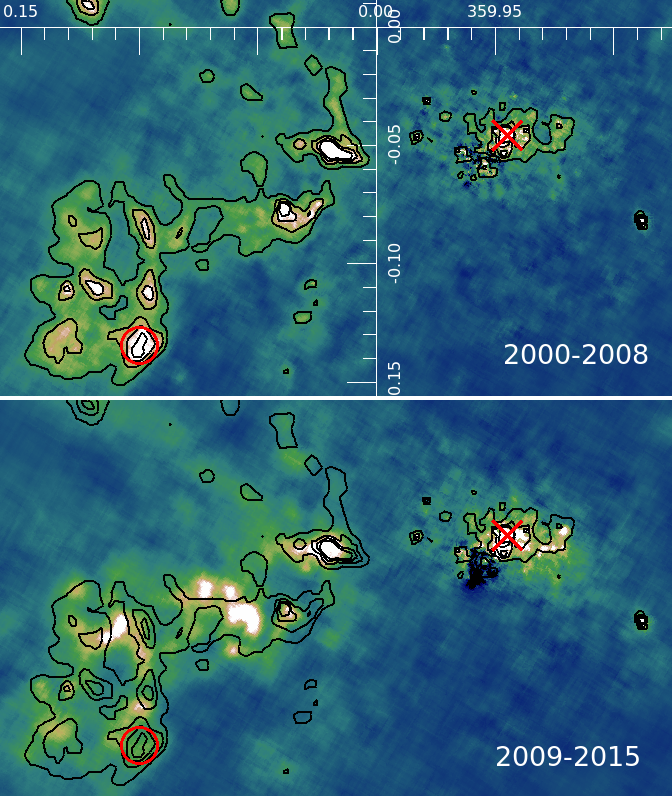}
\end{minipage}
\begin{minipage}{0.49\textwidth}
\includegraphics[trim= 0cm 4cm 0mm 2cm,width=1\textwidth,clip=t,angle=0.,scale=0.98]{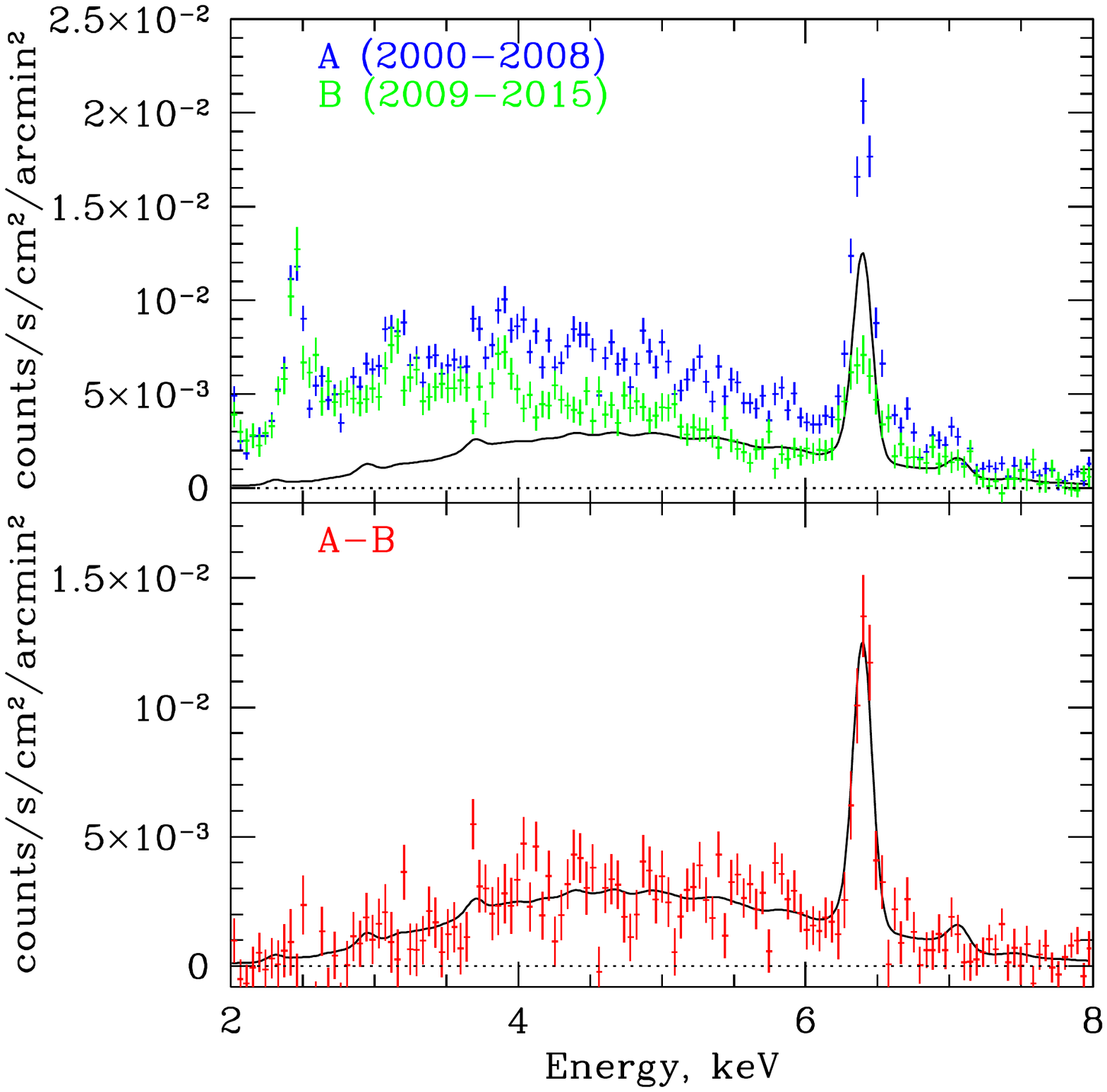}
\end{minipage}
\caption{Spectrum of a reflection component ``cleaned'' for the
  background contribution. {\bf Left:} \Chandra maps of the scattered
  emission in Sgr~A$^*$ region [modified version of the figure from
    \citet{churazov16}]. Images are in Galactic coordinates (see axis,
  $(l,b)$ in degrees). The top and bottom images show the maps
  averaged over 2000-2008 and 2009-2015 data, respectively. Contours
  on both images correspond to the top image to emphasize the
  difference in the maps. The red circles shows the region used to
  extract the spectra, A and B for the top and bottom images,
  respectively. In the 2000-2008 data, the reflected emission from this region was
  significantly brighter. {\bf Top-right:} The A and B spectra, extracted from the region shown with red circle, are shown with blue and green colours, respectively. The A spectrum is clearly brighter than the B spectrum. {\bf
    Bottom-right:} The difference between A and B spectra (red
  crosses). The difference spectrum removes the constant diffuse emission leaving the ``pure'' reflected component. The
  black line shows the best-fitting model of the reflected spectrum.
\label{fig:spec_ab}
}
\end{figure*}

Even a more convincing and powerful spectral test on the nature of the
``reflected'' component is possible for regions where we see
significant variability of the diffuse emission. This is
illustrated in Fig.~\ref{fig:spec_ab}, where the difference between
the spectra labelled as A and B, obtained during two different epochs,
are shown. For this exercise, the \Chandra data in 2000-2008 and
2009-2015 were used. We extracted the spectra from the red circle in Fig.~\ref{fig:spec_ab}, where variability of the diffuse emission
flux was particularly strong. In the difference spectrum, all steady
components must be gone and we expect to see a pure reflection
component. This procedure effectively simplifies the spectral
modeling and therefore can be used to place stronger constraints on the
reflected component alone. Indeed, the difference spectrum can be well
described by a pure reflection model with $\tau_T\sim 0.02$, cosine of
the scattering angle $\mu\sim 0$ and an additional hydrogen column
density of absorbing gas $N_H\sim 1.7~10^{23}~{\rm
  cm^{-2}}$. Potentially this approach provides the most robust way to
spectrally determine the position of the cloud along the line of
sight, since the shape of the spectrum, e.g., the equivalent width of
the 6.4 keV line, depends on the scattering angle due to the
difference in the angular distributions for scattered and fluorescent
photons \citep[see, e.g.,][]{1998MNRAS.297.1279S,2012A&A...545A..35C}
avoiding the necessity of modelling complicated
background/foreground. However, there are several caveats. One caveat
is the abundance of iron (and heavy elements in general), and as a consequence, the expected equivalent
width of the line is uncertain. The second is the geometry of the cloud that is
important in the case of substantial optical depth of the cloud. The
last uncertainty is purely instrumental and applies to the case shown
in Fig.~\ref{fig:spec_ab} - we used both ACIS-I and ACIS-S for this
analysis, correcting for the differences in the effective area, but
neglecting the difference in energy resolution, which is clearly an
oversimplification. This is the reason why we quote above the
approximate values of the parameters without uncertainties. However,
neither of these caveats is crucial and could be overcome to improve
such analysis. If we ignore these complications, e.g., assume that the
abundance used in the model is correct, then the minimal $\chi^2$ is
achieved for $\mu=-0.3$ (scattering angle $\sim$110\deg\!) and, at
$\sim$90\% confidence, $|\mu|<0.4$. This
implies that the cloud is located $\sim 8$~pc further away from us than Sgr~A$^*$. 
We note that this result agrees well with
the position estimate obtained by
\citet{churazov16} using the completely independent technique, although uncertainties in both approaches are large.

\section{Single-scattering approximation and time variability}
\label{sec:single}
In the previous section we have used detailed Monte Carlo simulations (including multiple scatterings)
of an isolated spherical cloud illuminated by a distant steady primary
source. We now would like to allow for a more elaborate geometry of
the gas distribution and include variability of the primary source.

\begin{figure}
\begin{minipage}{0.49\textwidth}
\includegraphics[trim= 1mm 0cm 0mm 2cm, width=1\textwidth,clip=t,angle=0.,scale=0.9]{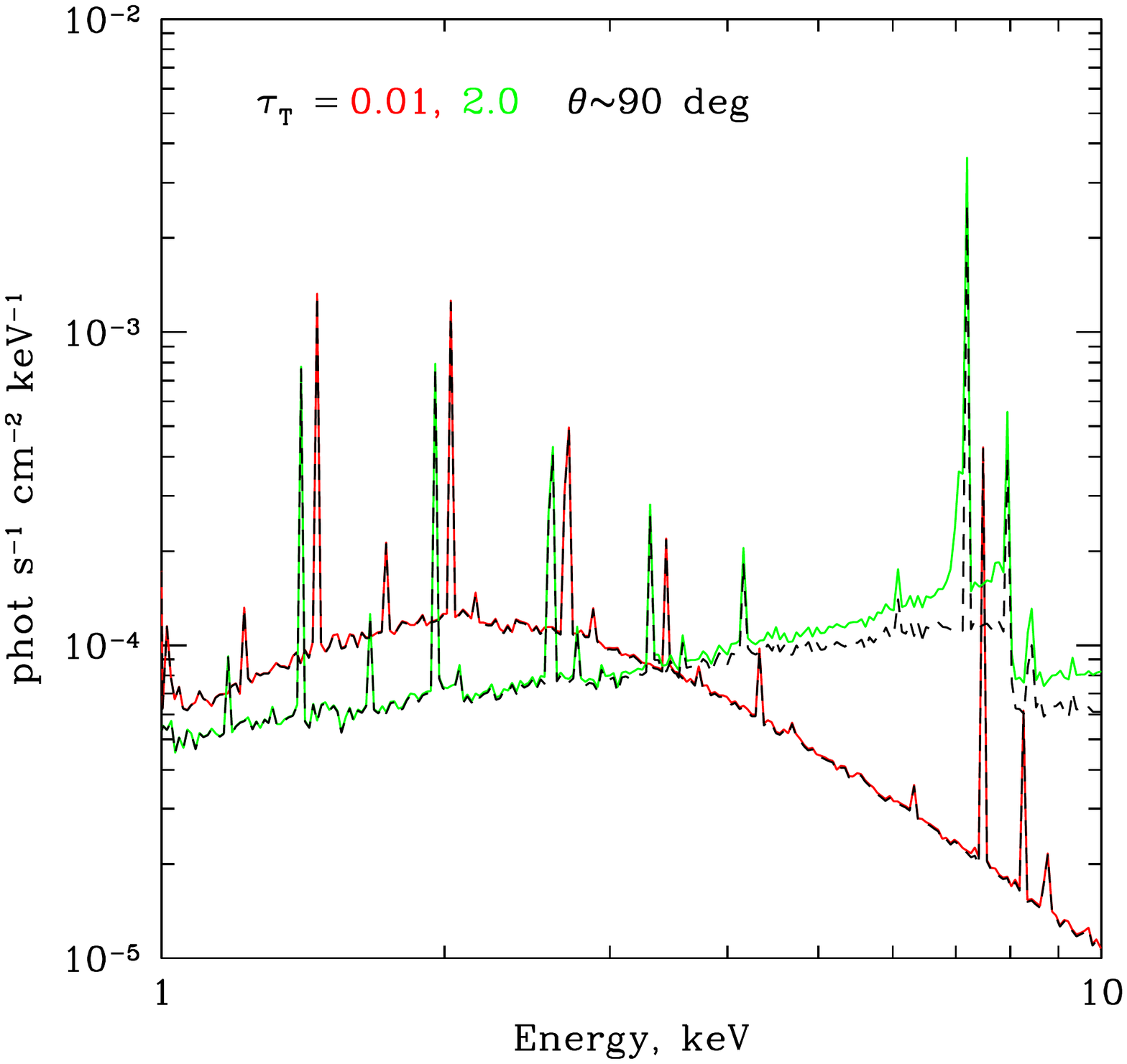}
\end{minipage}
\caption{Comparison of a single-scattering approximation with a full
  Monte Carlo treatment. The red and green curves correspond to the
  same spectra as shown in the left panel of Fig.~\ref{fig:spec_ma}
  for $\tau_T=0.01$ (red) and $\tau_T=2$ (green) cases. The black
  dashed lines show the contribution of the single-scattering
  component to these spectra. Clearly, the single-scattering
  approximation works well in both limits of small and large optical
  depths, except for the Compton shoulder of fluorescent lines (on the
  low-energy side of each line). As in Fig.~\ref{fig:spec_ma}, for
  clarity the plotted spectra are slightly shifted in energy.
\label{fig:spec_single}
}
\end{figure}

As the first step we note that for neutral gas with solar (or
super-solar) abundance of heavy elements, the photoelectric absorption
cross section is larger than the scattering cross section at energies
below $\sim$10~keV. This means that the role of multiple scatterings
is subdominal in this energy range for any value of the optical
depth. Indeed, in the limiting case of a small optical depth, the
probability of an additional scattering is simply proportional to
$\tau_T\ll 1$. In the opposite limit of an optical thick medium, the
probability of the scattering is set by the ratio of the scattering
cross section $\sim \sigma_T$ to the total cross section $\sigma \sim
\sigma_T +\sigma_{ph}(E)$, where $\sigma_{ph}(E)$ is the photoelectric
absorption cross section. Since this ratio is small (for energies
lower than $\sim$10~keV), the probability of an additional scattering
is also small.  As a simple illustration we show in
Fig.~\ref{fig:spec_single} several spectra calculated in
\S\ref{sec:modela} (see, Fig.~\ref{fig:spec_ma}, left panel) and the
contribution to these spectra from the photons, scattered only once
(dashed lines). It is clear that except for the Compton shoulder to
the left from the iron fluorescent line\footnote{The shoulder is
  absent in the single scattering approximation since the fluorescent
  line is formed only after the primary photons are absorbed.}, the
single-scattering approximation works well over the entire 2-10 keV
range. The calculation of the reflected spectra, their variability and
polarisation properties is straightforward in the single-scattering
approximation. Below we use this approximation to compute maps and
variability of the reflected component in the entire GC region.

\subsection{Distribution of molecular gas}
\label{sec:gas3d}

To simulate the properties of the reflected emission in the GC region, we
need a model of the 3D distribution of the gas in this region. Large
sets of high-quality sub-mm, IR and molecular data are available for the
central 1\deg \! region around the Galactic Center owing to its
extensive surveying in recent years \citep{Dahmen1998,Tsuboi1999,Pierce2000,Stolovy2006,Molinari2011,Jones2012}. 
Still, these data are capable of providing position-position-velocity (PPV)
information at best, so one has to assume some PPV vs. 3D position
correspondence in order to fully reconstruct 3D distribution of the
dense gas. Several models have been put forward to describe the
observed (rather complicated) PPV-distribution of the gas in the
region of interest in a concise and physically motivated way, such as
the 'twisted ring'-model proposed by \cite{Molinari2011} on base of
the \textit{Herschel} Hi-GAL survey, or the orbital solution by
\cite{Kruijssen2015} [for a thorough comparison of these models with
  the data provided by various gas tracers see \cite{Henshaw2016}].
Here we adopt the model of \citet{Kruijssen2015} as a baseline. Although
all such models are perhaps too simplistic for comprehensive
description of the gas distribution, it is well-suited for our
purpose, namely to get characteristic values of the expected reflected
component fluxes and polarization degrees over the central 1\deg \!
region around the Galactic Center.

In order to estimate the actual gas density along the path described
by the orbital solution of \cite{Kruijssen2015}, this solution was
used in conjunction with the CS (1-0) data by \cite{Tsuboi1999}.
Namely, we extracted the cells from the \cite{Tsuboi1999} data cube,
which falls sufficiently close to the PPV trajectory of
\cite{Kruijssen2015}, i.e. that the maximum projected distance of a
cell to the trajectory is not more than 2 arcmin and the maximum
difference along the velocity axis is 10 km/s. Every cell extracted
in this way was treated as a spherical cloud with diameter
equal to the cell width in sky projection, whilst its total H$_ 2$
mass is calculated according to the standard conversion procedure
described in Section 4.2 of \citep{Tsuboi1999} (namely, the fractional
abundance of CS molecule is adopted at X(CS)=$ 1\times 10^{-8}$
level and the excitation temperature $ T_{ex}=50 $ K). 
In order to eliminate gridding effects, additional Gaussian
scatter of the cloud positions was applied.

On top of this, we separately added the Sgr B2 cloud, since the CS
emission as a density tracer becomes increasingly less accurate for
that massive molecular clouds as a result of self-attenuation
\citep{Tsuboi1999,Sato2000}. Nevertheless, detailed density
distribution inside the Sgr B2 cloud can be reconstructed by means of
less optically thick tracers along with radiative transfer simulations
\citep{Schmiedeke2016}.  For the purpose of the current study,
however, the finest sub-structures are not very important, since only
a small fraction of the total mass ($\sim$1\%) is associated with
them. Moreover, the reflected signal from them is additionally
weakened since they are typically embedded in larger shells of less
dense gas (see e.g. a schematic Fig.1 in
\citealt{Schmiedeke2016}). Hence, we implemented only low and moderate
density envelopes, and Sgr B2 North, Main and South cores as described
in the Table B.3 of \cite{Schmiedeke2016}. The total mass of the Sgr
B2 complex included in our calculation is $\approx 2.2 \times 10^6
M_{\odot} $, while the total mass of the molecular gas extracted along
the trajectory of \cite{Kruijssen2015} is $\approx 9.2 \times 10^6
M_{\odot} $. We note here that only a fraction ($\sim 30\%$) of the
total molecular gas in CMZ ($ \sim 2-6\times 10^7 M_{\odot}$,
\citealt{Dahmen1998,Tsuboi1999,Pierce2000,Molinari2011}) has been
accounted for by this procedure.  In the base-line model the Sgr~B2
cloud is placed at the same distance as Sgr~A$^*$. We consider
different line-of-sight positions of the clouds in \S\ref{sec:sap}.
  
\begin{figure*}
\includegraphics[trim= 0mm 0cm 0mm 0cm,
  width=1\textwidth,clip=t,angle=0.,scale=0.9]{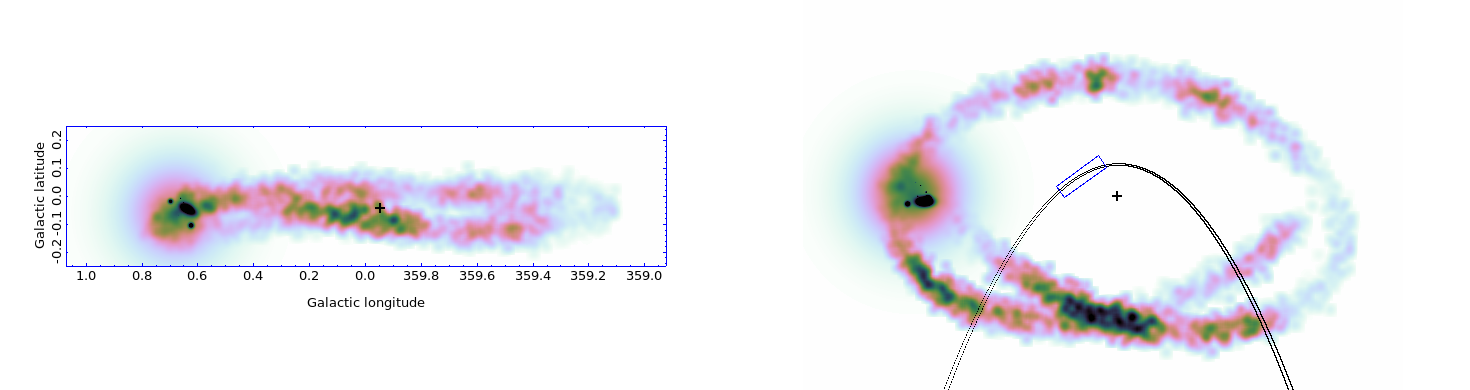}
\caption{Model of the distribution of the molecular gas in the
  $300\times 70 \times 300$~pc box (see \S\ref{sec:gas3d} for
  details). {\bf Left:} Projected gas column density onto $(l,b)$
  plane.  {\bf Right:} Projected gas column density onto $(l,z)$ plane
  (view from the top).  The large diffuse object in the left side of the plots
  corresponds to the Sgr B2 complex. In the base-line model the Sgr~B2
  cloud is placed at the same distance as Sgr~A$^*$. We consider
  different line-of-sight positions of the clouds in
  \S\ref{sec:sap}. The position of Sgr~A$^*$ is marked with the black
  cross. Thin black lines shows illiminated region for a model of a
  short outburts $\sim$110~yr ago \citep[see][]{churazov16}. The rwo lines
  correspond to the beginning and the end of a 5~yr long flare. The
  blue box shows the region where molecular gas should be located in
  the model of \citet{churazov16} to provide bright X-ray emission
  from the so-called ''Bridge'' region. There is a clear tension
  between the models, suggesting that either the gas distribution
  model is incomplete (see \S\ref{sec:gas3d}) or the X-ray emission
  from the ''Bridge'' is due to much more recent ($\sim$20~yr ago) or
  much older ($\sim$400~yr) flare that illuminates the front or the
  back parts of gas distribution shown in the figure. The degree of
  polarization of the reflected emission from the ''Bridge'' is
  expected to be much smaller in the 'recent' and 'old' flare
  scenarios (see \S\ref{sec:sap}), compared to the flare that happened
  $\sim$ 110~yr ago.
\label{fig:gas3d}
}
\end{figure*}

The resulting simulated distribution of the molecular gas is shown in
Fig.~\ref{fig:gas3d}. The left and
right panels show the projected column density of the gas onto $l,b$
and $l,z$ planes, respectively, where $l$ and $b$ are the Galactic coordinates and
$z$ is the distance along the line of sight. The simulated volume has
the size of $300\times 75 \times 300$~pc along $l$, $b$ and $z$ coordinates. The large diffuse object in the left side of the plots corresponds to the Sgr B2 complex placed at the same distance as Sgr~A$^*$.

This model certainly is a gross simplification of the true (unknown)
distribution of the gas. For instance, this model is completely devoid
from the gas in the innermost region around Sgr~A$^*$, while there is
clear evidence for the dense molecular clouds in this region, such as
the circumnuclear disc \citep{Becklin1982,Requena2012} and the 50 km
s$^{-1}$ cloud interacting with the Sgr A East
\citep{Coil2000,Ferriere2012}. Also, it does not include the molecular
Sgr~A/``Bridge'' complex (Fig.~\ref{fig:gas3d}), which is likely
located not further than $ \sim 25$ pc from Sgr~A$^*$
\citep{churazov16}. In future, the data on the degree of polarization
(\S\ref{sec:sap}) can differentiate between scenarios that differ in
the absolute value of the line-of-sight position $|z|$ of the
scattering cloud relative to the illuminating source. For now, the
model shown in Fig.~\ref{fig:gas3d} is suitable only for qualitative
characterization of the long-term evolution of the polarization and
flux of the reflected flux.

%%%%%%%%%%%%%%
%%%%%%%%%%%%%%%%%%%%

\subsection{Illuminating source}
We assume that the primary source is located at the position of
Sgr~A$^*$, namely, $(l_0,b_0,z_0)=(-8,-7,0)$~pc.  The light curve of the source is modelled as a single
outburst, that starts at time $t_1$ and ends at $t_2$, so that the
duration of the outburst is $t_{b}=t_2-t_1$. For our illustrative run
we have chosen $t_{b}=50~{\rm yr}$, $t_1=0$ and $t_2=50$, i.e. the
beginning of simulations at $t=0$ corresponds to the onset of the
outburst.  Motivated by the estimates of the flux from Sgr~A$^*$
required to explain the fluorescent emission from the Sgr~B2 clouds
\citep[see, e.g.,][]{1998MNRAS.297.1279S}, we set a fiducial 2-10 keV
luminosity of the source during the outburst to $L_X=10^{39}~{\rm
  erg~s^{-1}}$ with a power law spectrum with index $\Gamma=2$ \citep{2004A&A...425L..49R,2010ApJ...719..143T,2015ApJ...814...94M,2015ApJ...815..132Z}. As in \S\ref{sec:modela} the emission of the primary
source is assumed to be unpolarized.

\subsection{Absorption, Scattering, Polarization and Fluorescent Lines}
In the single scattering approximation approach the calculation of the observed
flux is straightforward. For a given cell $(l,b,z)$ of the 3D box, the
time delay $\Delta t$ from the primary source, Sgr A$^*$, to the scattering cell
and then to observer is
\begin{eqnarray}
  \Delta t =\left (\sqrt{(l-l_0)^2+(b-b_0)^2+(z-z_0)^2}-(z-z_0) \right ) /c
\end{eqnarray}
where $(l_0,b_0,z_0)$~pc are the coordinates of the primary source and
$c$ is the speed of light. Thus, the observer time $t$ translates into
the primary source time as $t_s=t-\Delta t$. If the value of $t_s$
falls in between $t_1$ and $t_2$, i.e., the time when the source is in
outburst, the flux $F(E)$ illuminating this cell is calculated as
\begin{eqnarray}
  F(E)=\frac{I(E)}{4\pi D^2} e^{-\tau_{1}(E)}, {\rm ~ where} \\
  \tau_1(E)=(\sigma_T+\sigma_{ph}(E))\int n(s) d s.
  \label{eq:tau}
\end{eqnarray}
In the above expression, $n$ is the gas density, $s$ is a coordinate
along the line connecting the primary source and the cell, $\tau_1(E)$
is the optical depth along this line and $D$ is the distance from the
primary source to the cell. Here $I(E)$ is the flux of the primary
source, such that $L_X=\int_2^{10} I(E)EdE$.

When calculating the scattered flux, including polarization, we use
Thomson differential cross section and store the polarized and the total flux
for each cell. Since we are using single-scattering approximation, the
orientation of the polarization direction is fully defined by the
relative positions of the primary source, the observer and the
scattering cell. In projection to the $l,b$ plane the polarization
direction is perpendicular to the line connecting $(l,b)$ position of
the cell and the primary source $(l_0,b_0)$. The procedure of
calculating the scattered flux is repeated for all energies.

The fluorescent lines of all astrophysically abundant elements are
included in the calculations and added to the corresponding bins of
the scattered spectrum. The fluorescent flux is emitted isotropically
and is not polarized.

The final step is the projection of the scattered flux onto $(l,b)$
plane, taking into account additional attenuation factor
$e^{-\tau_{2}(E)}$ due to the scattering and absorption along the line
connecting the cell and the observer\footnote{The calculations of
  $\tau_2$ and $\tau_1$ (the latter appears in equation~\ref{eq:tau}) are done separately,
  since for fluorescent line the energy of the absorbed photon changes
  to the energy of the fluorescent photon. For the photons of the
  continuum the change of energy is ignored.}. Any additional
attenuation outside of the simulated box is ignored in the
calculations, but can be easily added as a multiplicative factor to
the calculated spectrum.

\section{Time evolution, spectra and polarization based on the 3D model}
\label{sec:time}
\subsection{Time evolution}
\begin{figure*}
\includegraphics[trim= 0mm 0cm 0mm 0cm, width=1\textwidth,clip=t,angle=0.,scale=0.90]{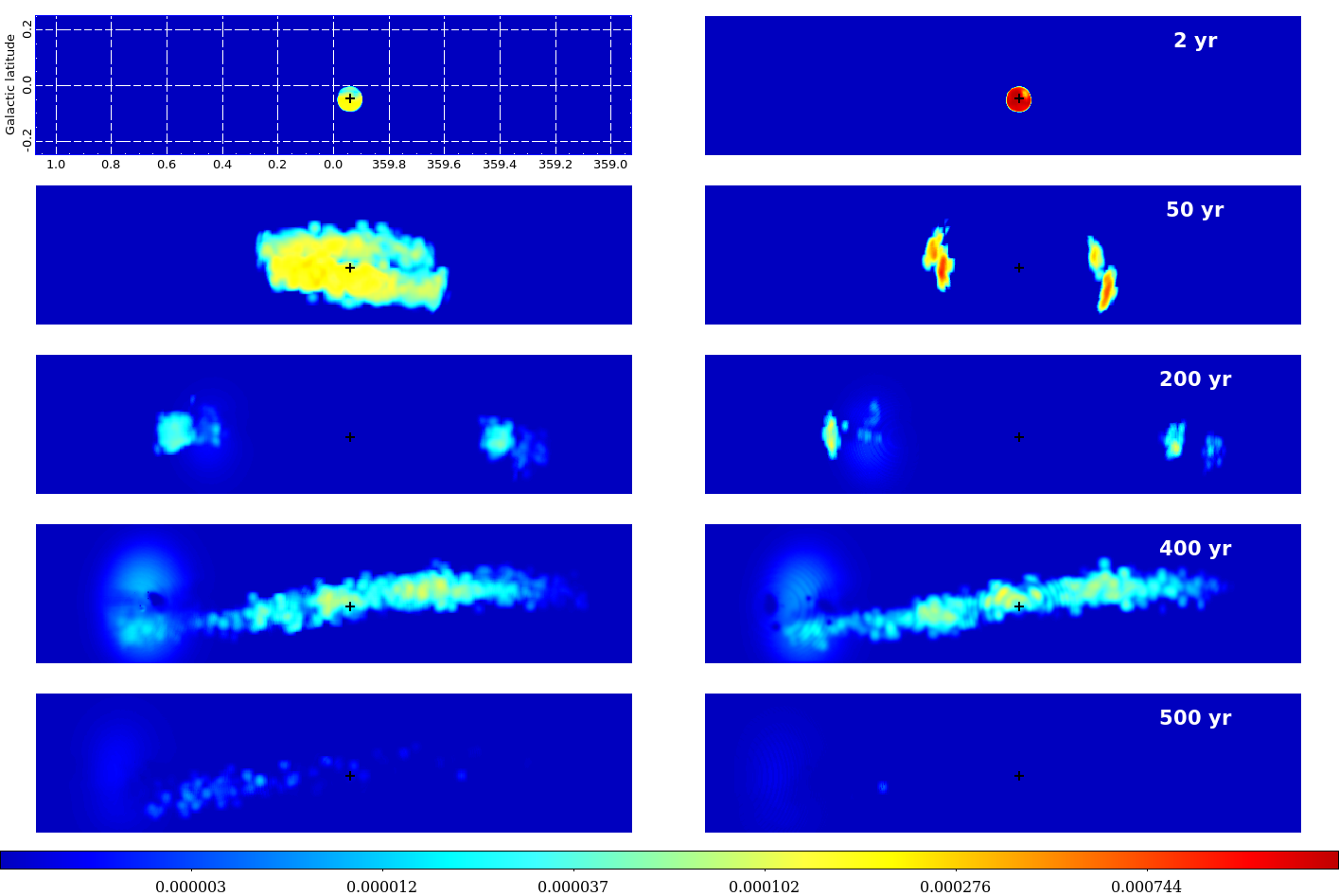}
\caption{Maps of the reflected component flux as a function of time
  for a single outburst of Sgr~A$^*$ (position of Sgr~A$^*$ is marked
  with the cross).  The left panel corresponds to a 50~yr long
  outburst, while the right panel shows the case of a shorter 5~yr
  long outburst.  The luminosity of the central source is 10 times
  larger for the 5~yr outburst, to provide the same fluence as for the
  longer outburst. The maps shown for moments $t=$2, 50, 200, 400, 500
  yr (from top to bottom; indicated in the right panel) since the
  onset of the outburst. The gas density distribution is described in
  \S\ref{sec:gas3d} and shown in Fig.~\ref{fig:gas3d}. As time
  progresses, the reflected signal first ``spreads'' to the sides from
  the primary source (2, 50, 200 yr frames) and then lights-up (frame
  400 yr) the more distant part of the molecular arm (see
  Fig.~\ref{fig:gas3d}) and the Sgr~B2 cloud. 500 yr after the
  outburst the photons leave the simulated region and the reflected
  component fades away.  Dark lanes across Sgr~B2 cloud, seen in the
  400 yr frame, are caused by the scattering/absorption along the path
  of the photons going from the primary source to the scattering cell
  and then to the observer. Naturally, any complicated time evolution
  of the primary source flux would lead
  to a distribution of the reflected emission that can be described as
  a linear combination of images, corresponding to short flares
  happening at different times. For instance, a sum of all five images
  in the right column would describe the expected outcome of five
  equal flares that happened 2, 50, 200, 400 and 500 years ago.
\label{fig:time}
}
\end{figure*}

We illustrate the expected time evolution of the reflected emission
produced by a 50 yr long outburst in Fig.~\ref{fig:time} (left panel) \footnote{ A movie showing the evolution of the 50~yr long
    outburst over $\sim550$~yr period is available at
    \verb!http://www.mpa-garching.mpg.de/!
    \verb!~churazov/gc_flare_v2.mp4!. In the movie,
the predicted reflected flux map was added to the observed \XMM 4-8 keV map.}. Initially,
the outburst illuminates the ``near'' side of the molecular gas distribution, which corresponds to the crossing streams in front of Sgr~A$^*$ (see the right panel of
Fig.~\ref{fig:gas3d}). The illuminated region quickly expands
($\sim$50~yr) since for the gas lying in front of the primary source,
the apparent velocity of motion is very high \citep[see,
  e.g.,][]{1998MNRAS.297.1279S}.  The flare then moves to the sides
(200~yr), then it  illuminates the more distant side of the trajectory, lying behind Sgr~A$^*$ (400~yr) and finally it fades away (500~yr).

For comparison, Fig.~\ref{fig:time} (right panel) shows the evolution for a shorter
outburst with $t_b=5$~yr. For this run the luminosity was increased by
a factor of 10, i.e. $L_X=10^{40}~{\rm erg~s^{-1}}$. Thus, the total
amount of energy released by the outburst $L_X t_b$ was the same as
for the longer outburst. In general, both scenarios share many
morphological features, although there are differences caused by the
different values of $t_b$. In particular, during the early phase the
bright spots are more compact (see, e.g., a snapshot at 50~yr after the onset
of the outburst). Also, for compact clouds, which have their entire
volume illuminated by a flare, the surface brightness is sensitive to
the luminosity of the outburst rather than to the fluence. As the
result, such clouds appear brighter in our $t_b=5$~yr scenario. In
contrast, for larger clouds the surface brightness is more sensitive
to the fluence, and they appear equally bright in both scenarios.

A similar exercise of modelling the long-term evolution of the
reflected emission was done by \citet{2014sf2a.conf...85C}, using the
twisted ring model of \citet{Molinari2011}. Qualitatively the
predictions are similar, implying that different patches within 100~pc
from the GC should brighten or fade away over hundreds of years. In
\citet{churazov16} we argued that it is possible to determine the
location of molecular complexes along the line of sight by comparing
the properties of spatial and time variations of the signal. The
estimates based on this approach suggest that the Fe K$ \alpha $ bright part of the Sgr~A complex (see
Fig.~\ref{fig:xmm_maps}) is located $\sim$10~pc further away from
Sgr~A$^*$. This is not consistent with the 3D gas distributions in the
models of \citet{Molinari2011} or \citet{Kruijssen2015}. This is not
surprizing, given that significant fraction of the gas is not
accounted for by the model adopted here (see \S\ref{sec:gas3d}).
Therefore, more work is needed to unambiguously position molecular
clouds in 3D.  However, it is clear that once few clouds are
accurately positioned and the light curve properties of the primary
source are determined (presumably via observations of several very
localized clumps) the true power of the X-ray tomography of the
molecular gas can be employed. One of the possible way to determine
the position of a cloud along the line of sight is to measure the
degree of polarization, discussed in the next subsection.

\subsection{Polarization maps}
\begin{figure}
\includegraphics[trim= 0mm 0cm 0mm 0cm, width=1\textwidth,clip=t,angle=0.,scale=0.49]{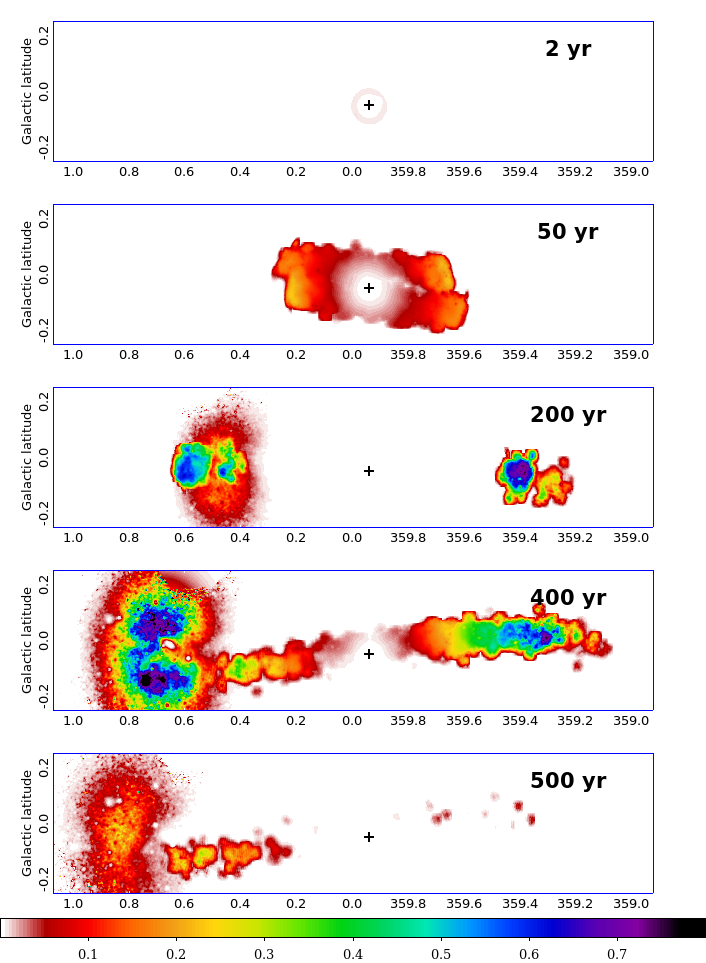}
\caption{Expected degree of polarization for the continuum in the 4-8
  keV band (excluding fluorescent lines), obtained by combining the
  observed X-ray 4-8 keV image (see top panel in
  Fig.~\ref{fig:xmm_maps}) with the simulated maps (see
  Fig.~\ref{fig:time}). The colorbar at the bottom of the plot
  shows the degree of polarization.  The 50~yr long outburst model is
  used to simulate the reflected emission. All five panels correspond to
  the same moments (2, 50, 200, 400, 500~yr) as in
  Fig.~\ref{fig:time}. For the central regions the degree of
  polarization is always low, since the scattering angle is
  either small or close to 180\deg (given the 3D geometry shown in
  Fig.~\ref{fig:gas3d}). The largest degree of polarization is
  expected for the Sgr~B2 cloud, since in these simulations it is
  located at the same distance as Sgr~A$^*$.
\label{fig:polmap}
}
\end{figure}

The calculation of polarization in the single-scattering approximation
is straightforward \cite[see also][ for the discussion of polarized
  flux after muliple scatterings]{2002MNRAS.330..817C}, since the
direction  and the degree of polarization depend only
on the scattering angle. Of course in real observations an additional,
presumably unpolarized, background/foreground emission should be
present. One can use existing \XMM images (see
Fig.~\ref{fig:xmm_maps} and \citealt{2015MNRAS.453..172P}) to approximately account for the contribution
of this emission by assuming that the observed emission is
unpolarized\footnote{In fact some of the observed emission in X-ray
  images is due to scattered radiation an can therefore be polarized.}
and adding this unpolarized signal to the simulated emission of the
scattered component. In Fig.~\ref{fig:polmap} we show the expected
polarization of the continuum emission in the 4-8 keV band obtained by
combining simulated and observed images.
The five panels shown correspond to the same moments (2, 50, 200, 400,
500~yr) as in Fig.~\ref{fig:time}. For the central regions the degree
of polarization is always low, since there the scattering angle is
either small or close to 180\deg\!  (given the 3D geometry shown in
Fig.~\ref{fig:gas3d}). The largest degree of polarization is expected
for the Sgr~B2 cloud, since in these simulations it is located at the
same distance as Sgr~A$^*$. Note also, that the degree of polarization
shown in Fig.~\ref{fig:polmap} depends both on the intrinsic degree of
polarization $p_{int}$, set by the scattering angle, and the relation
between the intensity of reflected $I_R$ emission relative to the
unpolarized background $I_B$, i.e. $p=p_{int}/(1+I_B/I_R)$. The
reflected intensity depends on the luminosity of the illuminating
source and the duration of the outburst, while the intrinsic
polarization depends only on the geometry of the problem.

\subsection{Spectra and polarization}
\label{sec:sap}
\begin{figure*}
\begin{minipage}{0.49\textwidth}
\includegraphics[trim= 1mm 0cm 0mm 2cm, width=1\textwidth,clip=t,angle=0.,scale=0.9]{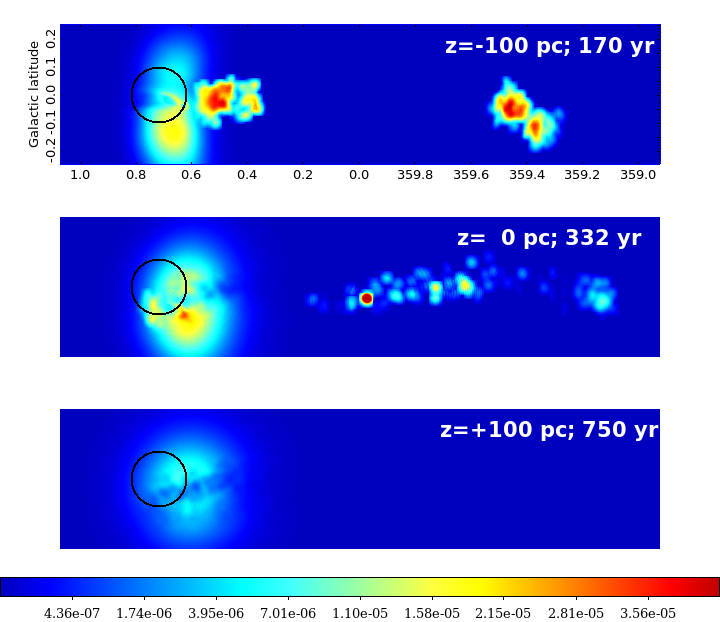}
\end{minipage}
\begin{minipage}{0.49\textwidth}
\includegraphics[trim= 1cm 4cm 0mm 2cm,width=1\textwidth,clip=t,angle=0.,scale=0.9]{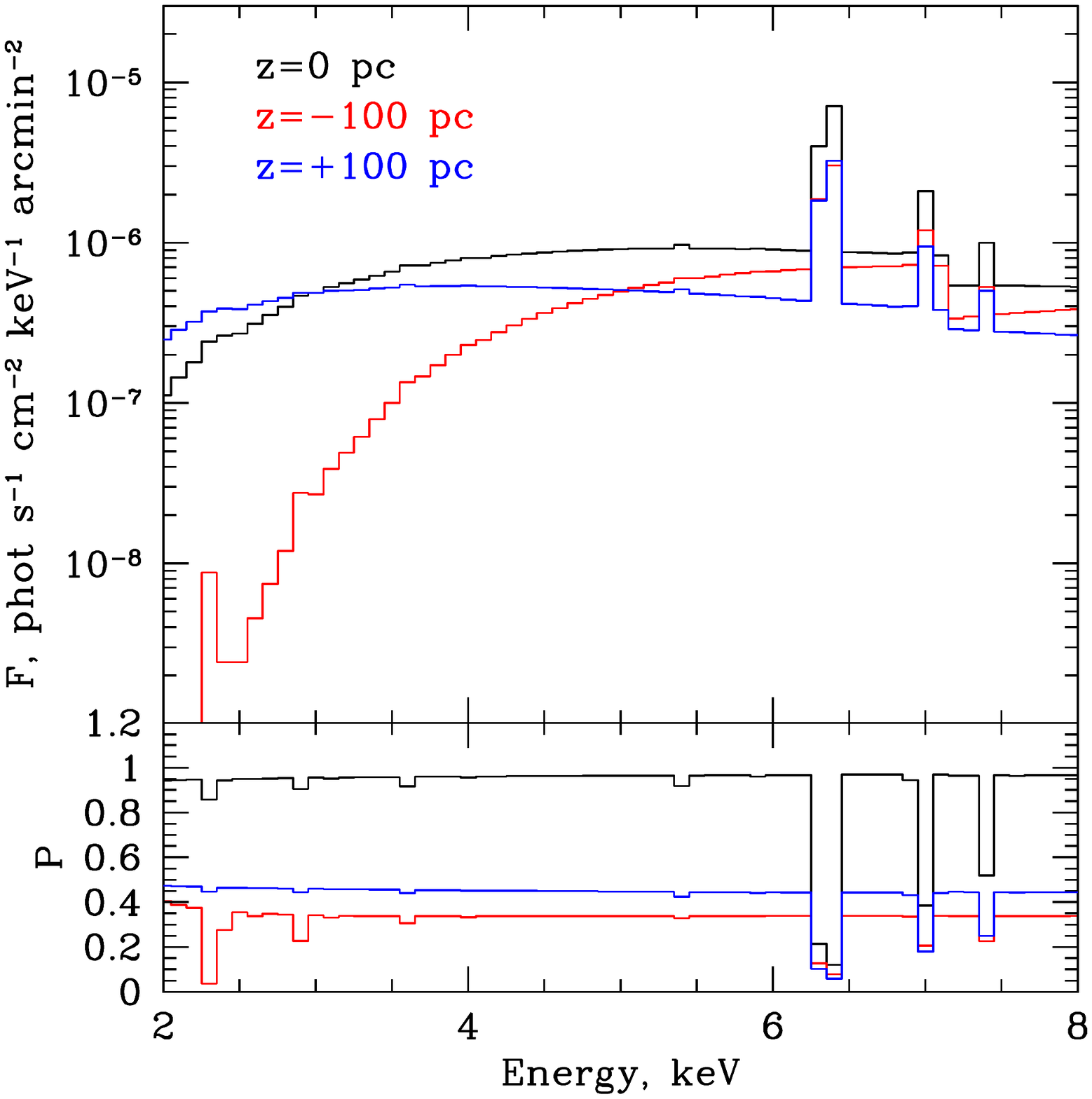}
\end{minipage}
\caption{Expected spectra for different positions of Sgr~B2 along the
  line of sight. {\bf Left:} Maps of the scattered emission in the 4-8
  keV band. The clouds is placed at -100, 0, +100 pc, relative to Sgr
  A* (from top to bottom). The time of the observation is 170, 332,
  750 yr after the outburst for these three panels. {\bf Top-right:}
  The spectra of the reflected component, corresponding to three
  panels on the left are shown by the red (-100~pc), black (0~pc) and
  and blue (+100~pc) lines. {\bf Bottom-right:} The degree of
  polarization for the scattered component. As expected, the degree of
  polarization is close to 100\% for the case when the cloud is at the
  same distance as the primary source (the black line), but drops
  significantly if the centre of the clouds is shifted along the line
  of sight by $\pm 100$~pc (red and blue lines).
\label{fig:spec_mb}
}
\end{figure*}

A more accurate estimate of the reflected emission and the degree of
polarization can be easily obtained in the single-scattering
approximation by calculating the expected spectra. For this purpose, we
used a 15 pc (radius) circle centered at the Sgr B2 position and
calculated the total spectrum coming from this region. We then
repeated the same exercise varying the position of the centre of the Sgr B2 cloud
along the line of sight ($z=-100, 0, 100$~pc). The resulting spectra
are shown in the right panel of Fig.~\ref{fig:spec_mb}. The left panel of
the figure shows the snapshots corresponding to the moments when the cloud
is illuminated by the outburst.
As expected the degree of polarization changes with the scattering
angle as $\displaystyle p=(1-\mu^2)/(1+\mu^2)$, i.e. it is close to
100\% for the case when the cloud is at the same distance as the
primary source (the black line), i.e., $p\sim 1$. Shifting the cloud
by $\pm 100$~pc, implies $|\mu| \sim 0.7$ and $p\sim 35$\%. The actual
values of polarization are slightly larger (blue and red lines) than
this value because the angular size of the cloud (as seen from the Sgr
A$^*$ position) is large and for the selected snapshots the effective
scattering angle is slightly larger. As expected \citep[see,
  e.g.,][]{2002MNRAS.330..817C}, the degree of polarization drops at
the energies of prominent fluorescent lines. Note, that the degree of
polarization in Fig.~\ref{fig:spec_mb} is computed for pure reflected
component. As we have shown in \S\ref{sec:comp} the contribution of
the unpolarized continuum in the regions with prominent reflected
components is $\sim$50\%. Therefore, we expect the observed polarization
to be factor of $\sim$2 smaller.

\section{Prospects for \XIPE}
\label{sec:xipe}
\subsection{Feasibility}
The aim of the polarization measurements is three-fold:
\begin{itemize}
\item Establish a non-zero polarized signal of the continuum emission
  as a proof that it is due to scattered emission;
\item Find the polarization angle to identify the location of the
  primary source;
  \item Measure the degree of polarization to estimate the
    line-of-sight location of the scatterer relative to the primary
    source\footnote{Note, that the degree of polarization depends on the cosine of the scattering angle, i.e. only the absolute value of the line-of-sight position $|z|$ can be estimated. The spectral and variability data can help to lift this degeneracy.}.
\end{itemize}
In this section we evaluate the ability of the \XIPE mission
\citep{2013ExA....36..523S} to detect polarized reflected emission from the GC molecular clouds. Below we use the effective area and the
modulation factor from \citet{2013ExA....36..523S}. A similar question
has already been addressed in \citet{2015A&A...576A..19M} based on
the assumed positions of several clouds.  In our simulations the
uncertainties in the choice of the 3D gas distribution and the
parameters of the outburst are considerable. Moreover, we believe that
some of the regions that are bright in the reflected component, are
not adequately described by this model \citep[see][]{churazov16}.  We
therefore consider simulations as a guideline, but use actual
\XMM observations to estimate the fluxes.

From the observed and simulated spectra (see Figs.~\ref{fig:spec_ma},
\ref{fig:xmm_maps}, \ref{fig:spec_mb}) it is clear that at low
energies the photoelectric absorption can suppress the reflected
component if a cloud is illuminated from the back. At the same time,
above $\sim$6~keV the spectrum is dominated by the 6.4 keV from
neutral iron and 6.7-6.9 keV lines from the hot plasma. Since we
expect the emission in these lines to be unpolarized, it does not make
sense to use the energy range 6-7.5 keV when searching for polarized
emission\footnote{At the same time, a drop of the polarization degree
  at the energies of fluorescent or plasma emission lines is a useful
  testable prediction of the model.}. Thus, we have selected an energy
band from 3 to 6 keV for the feasibility study. The problem of
evaluating the degree of polarization and the polarization angle
was discussed recently by
\citet{2010SPIE.7732E..0EW,2013ApJ...773..103S,2015ApJ...801...21M}. Here
we adopt the simplest approach possible. To estimate the expected S/N,
we have extracted spectra from three $4'$ (radius) circles from the
Sgr~B2, the ``Bridge'' and Sgr C regions. For this purpose all
publicly available \XMM (MOS) data were used. Only the data in the
3-6 keV band, where continuum emission dominates, were used for the
sensitivity estimates. As mentioned in \S\ref{sec:comp} it is
reasonable to assume that in these regions about half of the signal is
due to the reflected component, while another half is due to
(unpolarized) background/foreground emission.

We assume that for a source with a polarization angle $\phi_0$ the
signal recorded by a polarization-sensitive detector is described by
the following expression
\begin{eqnarray}
  D(\theta)=[B+S]+[S p m]\cos 2\left (\theta - \phi_0 \right ) + \eta,
  \label{eq:d}
\end{eqnarray}
where $\theta$ is the phase, $B$ is the intensity of a steady
(unpolarized) background, $S$ is the sources flux, $p$ degree of
polarization of the source flux, $m$ is the known modulation factor in
this energy band and $\eta$ is the noise that we assume, for
simplicity, to be Gaussian and the same for all phases. We further
assume that the background is known and set $B=0$.

This signal can be approximated with a model
\begin{eqnarray}
  M_1(\theta)=C_1+C_2\cos 2\left (\theta - \phi \right ),
  \label{eq:mop1}
\end{eqnarray}
where $C_1$, $C_2$ and $\phi$ are the parameters of the model. Even
more convenient is the parametrization
\begin{eqnarray}
  M_2(\theta)=A_1+A_2\cos 2(\theta)+A_3\sin 2(\theta),
  \label{eq:mop2}
\end{eqnarray}
where $A_1$, $A_2$ and $A_3$ are free parameters.  The best-fitting
values can be determined via simple $\chi^2$ minimization,
\begin{eqnarray}
\chi^2=\Sigma_\theta \left (
\frac{M(\theta)-D(\theta)}{\sigma}\right )^2\rightarrow {\rm min},
\label{eq:chi2}
\end{eqnarray}
where $\sigma^2=Var(\eta)$ is the variance of the noise $\eta$. The advantage of the second model is
that the explicit expressions of the best-fitting parameters $A_1$,
$A_2$ and $A_3$ can be immediately written as
\begin{eqnarray}
\sum_\theta
D(\theta)f(\theta)/\sum_\theta f^2(\theta),
\end{eqnarray}
where $f(\theta)=1$,
$\cos (2\theta$) and $\sin (2\theta)$ for $A_1$, $A_2$ and $A_3$,
respectively.  The values of $A_1$, $A_2$ and $A_3$ have Gaussian
distributions with standard deviations
$\sigma_{A_1}=\sigma/\sqrt{n_\theta}$,
$\sigma_{A_2}=\sigma_{A_3}=\sqrt{2}~\sigma_{A_1}$. Note, that
functions $f$ for different $A$'s are mutually orthogonal, i.e.,
resulting contributions of the noise to $A_1$, $A_2$ and $A_3$ are
uncorrelated.  The number of bins $n_\theta$ over $\theta$ should be
sufficiently large to properly sample the model. Even if the actual
number of counts per bin is small, but the total number of counts is
$\gg 1$, the $\chi^2$ minimization can still be used
\citep[][]{1996ApJ...471..673C} to get an unbiased estimate of the
best-fitting parameters and correct confidence regions.

\subsection{Known polarization angle}
If the polarization angle $\phi_0$ is known (e.g., when the location
of a primary source is known) the minimization for the model given by
equation~(\ref{eq:mop1}) reduces to a linear problem and explicit expression
for the coefficients $C_1$ and $C_2$ can be easily written. In the
frame of this model the expected significance of the polarized signal
detection $S_{p}$ is simply
\begin{eqnarray}
  S_{p}=\frac{pm}{\sqrt{2}}\frac{S\sqrt{n_\theta}}{\sigma}=\frac{pm}{\sqrt{2}} S_{C_1}
  \label{eq:snp}
\end{eqnarray}
where $S_{C_1}\equiv S_{A_1}=S\sqrt{n_\theta}/\sigma$ is the expected
signal-to-noise ratio for $C_1\equiv A_1$ value, i.e., the
significance of the total source flux detection. This expression can
be easily generalized for an arbitrary spectral shape and energy
dependent modulation factor and polarization.

\subsection{Null hypothesis of zero polarization}
We can now relax the assumption that the angle is known and use the
model given by equation~(\ref{eq:mop2}) to obtain best-fitting values of
$A_1$, $A_2$ and $A_3$. Consider first a case when true polarization
of the source is zero, i.e., $p=0$. The probability $P$ of finding a
value of $x=\sqrt{A_2^2+A_3^2}$ larger than a certain threshold $t$
is, obviously,
\begin{eqnarray}
P(x > t)= e^{-\frac{t^2}{4 \sigma^2_{A_1}}}.
\end{eqnarray}
The above expression can be used to determine the significance of the
polarization detection in observations. For the feasibility studies
the minimum detectable polarization $p_{min}$, given total source flux
detection significance $S_{A_1}$ and the probability of false
detection $P$ (e.g., $P=10^{-3}$), is simply
\begin{eqnarray}
p_{min}=\frac{1}{m}\frac{2}{S_{A_1}}\sqrt{-\log P}.
\end{eqnarray}

\subsection{Confidence regions}
We now outline the simplest procedure of estimating the confidence
regions from the observed data. Given the best-fitting values of
$A_1$, $A_2$ and $A_3$ the angle $\phi$ and polarization $p$ can be
evaluated as
\begin{eqnarray}
  \phi=\frac{1}{2}\arctan \left ( \frac{A_3}{A_2} \right ) \\
  p=\frac{\sqrt{A_2^2+A_3^2}}{A_1 m}=\frac{x}{A_1 m},
  \label{eq:phip}
\end{eqnarray}
where as before $x=\sqrt{A_2^2+A_3^2}$. The value of $\phi$ does not
depend explicitly on $A_1$, while the expression for $p$ includes
$A_1$ as a normalization factor. For simplicity, we consider below the
confidence regions for the parameters $\phi$ and $x$ that depend
only on the values of $A_2$ and $A_3$.

\subsubsection{Confidence region for a pair $(\phi,x)$.}
The easiest case is a confidence region\footnote{Here we follow a
  frequentist approach, where confidence region at a given
  confidence level $P$ (e.g., $P=0.95$) should include true values in
  the fraction $P$ of multiple repeated experiments.}  for the pair of
parameters $(\phi,x)$. Consider a difference in the values of $\chi^2$
(equation~\ref{eq:chi2}) for a model with a given set of parameters and for
the model with best-fitting parameters ($\chi^2_{bf}$):
\begin{eqnarray}
\Delta \chi^2 =\chi^2-\chi^2_{bf}.
\end{eqnarray}
Since the distribution of $A_2$ and $A_3$ around true values has a
bivariate Gaussian distribution, the confidence region in the
$(A_2,A_3)$ space can be calculated as the 2D region where $\Delta
\chi^2$ is less or equal to a threshold value $\Delta_2$, such that
the probability of $\chi^2_2<\Delta_2$ is equal to $P$. Here
$\chi^2_2$ is a $\chi^2$ distribution with two degrees of freedom,
i.e., for $P=0.9$, $\Delta_2=4.61$. The corresponding region in the
$(\phi,x)$ will obviously enclose true values of $\phi$ and $x$ each
time when a pair of true values $(A_2,A_3)_{true}$ is enclosed.

\subsubsection{Confidence region for $\phi$ as a single parameter of interest.}

We are often interested in the confidence regions separately for
$\phi$ and $x$. For $\phi$ (as a single parameter of interest) the
confidence region can be evaluated by finding the minimum of the
$\chi^2$ for each value of $\phi$ (minimizing over $C_1$ and $C_2$ in
equation~\ref{eq:mop1}) and selecting the range of $\phi$ such that $\Delta
\chi^2$ is less or equal to a threshold value $\Delta_1$, such that
the probability of $\chi^2_1<\Delta_1$ is equal to $P$, where
$\chi^2_1$ is a $\chi^2$ distribution with one degree of freedom;
i.e., for $P=0.9$, $\Delta_1=2.71$. For feasibility studies, when the
polarization signal has high significance, i.e. $S_p\gg 1$, one can
derive a simple estimate of the expected error in $\phi$:
\begin{eqnarray}
  \phi_{err}=\frac{1}{2} \arcsin \left (\frac{\sqrt{\Delta_1}}{S_{p}}\right )\approx 28.65 \left (\frac{\sqrt{\Delta_1}}{S_{p}}\right )~~{[\rm deg]}.
  \label{eq:phierr}
\end{eqnarray}

\subsubsection{Confidence region for $x$ (and $p$) as a single parameter of interest.}

For the value of $x$ (as a single parameter of interest) the recipe of
using $\Delta \chi^2 < \Delta_1$ is not exact. This is clear, for
instance, in case if true polarization is zero or very small, i.e.,
$S_p\ll 1$. In this limit the model with the true value of
polarization is obviously insensitive to the value of $\phi$. In this
case the value of $\Delta \chi^2$ with respect to the best-fitting
model (see equation~\ref{eq:mop2}) has a $\chi^2$ distribution with two
degrees of freedom. However, when the significance of the polarization
increases, the accuracy of the confidence evaluation based on
$\Delta_1$ increases. For $S_p\gtrsim 1$ one can already safely use
this approach. In real observations the value of $S_p$ is not known,
but one can use in lieu the significance of non-zero detection of
polarized signal.

As the last step (see equation~\ref{eq:phip}), the value of $x$ has to be converted to polarization
$p=x/(m A_1)$. One can expect that in majority of real objects, the
total source flux $A_1$ will be determined with much higher accuracy
than the polarized component $x$. In this case the confidence
regions can simply be extended based on the relative uncertainty in
$A_1$. For most cases this should be an acceptable solution.

\subsection{Predictions for \XIPE}

For the three selected regions we know the flux level of the scattered
and background component from the \XMM data and have a
reasonable approximation of the spectrum shape in the 3-6 keV band
(see Fig.~\ref{fig:xmm_4reg}). Moreover, the angle $\phi$ is also
known, assuming that Sgr~A* is the primary source of X-ray radiation.
We also know the effective area and the modulation factor of the
instrument. What is unknown is the degree of polarization $p$ of the
scattered component. The value of $p$ depends solely on the shift $z$
along the line of sight of the scattering site relative to the primary
source. One natural assumption would be that $\displaystyle |z|\sim
\sqrt{(l-l_0)^2+(b-b_0)^2}$, i.e., the shift along the line of sight
is of the same order as the projected distance from the primary source
\citep[see, e.g.][]{churazov16}. This estimate suggests that
$0.35\lesssim p \lesssim 1$ (see \S\ref{sec:sap}). On the other hand, the 3D model of the
gas distribution adopted in \S\ref{sec:gas3d} suggests that in the gas
seen in projection close to Sgr~A$^*$ is located either in front or
behind the source with $\displaystyle |z|$ much larger than the
projected distance. This would imply much lower degree of
polarization.  In Table~\ref{tab:xipe} we quote our estimates of the
polarization component S/N ratio (see equation~\ref{eq:snp}) assuming
$p=0.5$ in the scattered component in all three region (the net
polarization with account for unpolarized background will, of course,
be factor $\sim 2$ lower). This estimate assumes that the $\phi$ is
known. Clearly the polarization can be easily detected even if the
intrinsic polarization is 5-10 times lower, corresponding to a scattering
angle of $\sim$20 (or 180-20=160) degrees. This scattering angle corresponds
to the shift along the line of sight $\sim$3 times larger than the
projected distance. This means that it is likely that polarization
will be significantly detected for any plausible distribution of
scattering clouds and will provide reliable determination of their line of
sight positions.

\begin{table*}
  \caption{Estimated detection significance of polarization from three
    selected regions within 100~pc from GC with \XIPE for 1 Msec
    exposure. Flux is the total observed flux by \XMM, i.e.,
    $S+B$ as in equation \ref{eq:d}. It is assumed that the
    contributions of $S$ and $B$ are equal (see \S\ref{sec:2t});
    $p=0.5$ is set for the scattered component $S$.
    \label{tab:xipe}
  }
  \begin{tabular}{|r | r | r | r | r }
    \hline
    Region & Flux, 3-6 keV & Detection significance, $S_p$ & Error on angle, $\phi_{err}$ \\
           & ${\rm erg~s~cm^{-2}}$ & (for $p=0.5$) & deg~~(for $p=0.5$) \\
    \hline
    Sgr~B2 & $2\times10^{-12}$  & $\sim$20 & $\sim 1.4$ \\
    Sgr A /``Bridge''   & $1\times10^{-11}$  & $\sim$50& $\sim 0.6$ \\
    Sgr C & $2\times10^{-12}$ & $\sim$20 & $\sim 1.4$ \\
    \hline
  \end{tabular}
\end{table*}

The above results broadly agree with the previous estimates
\citep{2015A&A...576A..19M} for \XIPE, although there are interesting
differences. In our case we used the observed X-ray spectra (rather than
simulated) for the feasibility study and we used a simple decomposition of
the spectra into reflected component and unpolarized one. This makes
the predictions insensitive to the modelling in the part related to
the intensity of the reflected emission. The main remaining  uncertainty 
is the cloud positions along the line of sight that sets the
polarization. Based on our previous study \citep{churazov16} of the variability of the
the Sgr~A/``Bridge'' complex, we believe that this gas is not strongly
shifted along the line of sight with respect to the Sgr~A*. This therefore
predicts high degree of polarization of its reflected emission and makes Sgr~A/``Bridge'' complex a promising (and, perhaps, even the most promising, given its brightness) target for \XIPE.

Given that the majority of the X-ray emitting clouds in the GC region
are variable on time scales of few years - decades \citep[see, e.g.,][]{2009PASJ...61S.241I,2010ApJ...714..732P,2013A&A...558A..32C,2015ApJ...814...94M} it is not possible to make
firm predictions on the expected fluxes for future missions more than
1-2 years ahead. Historic observations and our simulations show that
it is possible that another set of regions will be bright then. Unless
we are particularly unlucky, the flux level of those bright regions
will be similar to the flux levels we saw in the past 20-30 years. It
therefore makes sense to select targets during
the active phase of the mission (or $\sim$1-2 years before).

\section{Discussion}
\label{sec:discussion}
\subsection{Is Sgr~A* the primary source?}
In a addition to Sgr~A*, the molecular clouds in the Galactic Center region can be illuminated by  stellar mass X-ray
sources, namely X-ray binaries or flaring magnetars, located in the same
region \citep[see, e.g., relevant discussions in][]{2013ASSP...34..331P,2013ApJ...775L..34R,2013A&A...558A..32C}.

Clearly, any powerful source (with luminosity exceeding $10^{39}~{\rm erg\,s^{-1}}$) located in the close vicinity of the Galactic Center could mimic Sgr~A* flare in terms of the polarization pattern or the long-term variability of the reflected emission.   A short-term variability and spectral shape information can be used to distinguish between i) hard and very short ($\lesssim 1$ s) giant magnetar flare \citep[e.g.][]{Hurley2005}, ii) an outburst of a (currently
dim) ultraluminous X-ray source ($ L_{X}\sim 10^{40}$ erg s$ ^{-1} $) with
the spectrum typically characterized by a cut-off like curvature at $ \sim 5$ 
keV (e.g. \cite{Walton2014}), although the detection of the X-ray reflection signal by {\it INTEGRAL} \citep{2004A&A...425L..49R} and {\it Nustar} \citep{2015ApJ...815..132Z} in the 10-100 keV range argues against the strong cutoff in the spectrum iii) or a longer duration ($\sim 10 $ yr) and
smaller luminosity ($ L_{X}\sim 10^{39}$ erg s$ ^{-1} $) active state of a
more secular X-ray binary (e.g. similar to GRS 1915+105).

On the contrary, a bright source shifted from the projected position of Sgr~A* would produce different polarization and long-term variability patterns, which can be used to falsify the hypothesis that Sgr~A* is the primary source. The variability test works best if the molecular gas distribution is known. On the other hand, the orientation of the polarization plane is least sensitive to the details of the model, except for the mutual projected positions of the source and the cloud. The strongest constraints would come from those clouds (bright in reflected emission) that subtend the largest angle in the $(l,b)$ plane from the Sgr~A* position. This strengthens the case of a spatially resolved polarization measurements of the Sgr~A/''Bridge'' region, which is located close to Sgr~A* and subtends almost 70\deg (as seen from Sgr~A*).
     
A more complicated situation arises when multiple illuminating sources are present and/or intrinsic beaming and corresponding variability of the illuminating radiation. The predictions of the resulting reflected signal are discussed, e.g.,  in \cite{Molaro2014} and \cite{2016MNRAS.457.3963K}.

\subsection{Sgr~A* variability}
As we have shown in \S\ref{sec:time}, for the adopted gas distribution
model, the reflected emission is visible for $\sim 500$ yr after a flare of the illuminating source.  Hence,
observing reflection signal at different positions across this region
in principle allows one to reconstruct the Sgr~A* activity over past $
\sim 500$ yr.  Currently, the observations suggest, that $\sim 100 $ yr ago, Sgr~A*  had
an X-ray luminosity in excess of $ 10^{39}~{\rm erg~s^{-1}}$ for a period of few years
\citep[e.g.,][]{2004A&A...425L..49R,2010ApJ...719..143T,churazov16}, which is $ 10^6$ times brighter
than the characteristic Sgr~A*'s quiescent X-ray luminosity \citep[e.g.,][]{2003ApJ...591..891B,2013Sci...341..981W}. A
number of scenarios have been proposed to explain such behavior,
including accretion of a stellar material captured from the winds of 
nearby stars \citep{Cuadra2008} or supplied by their tidal interaction
\citep{Sazonov2012}; complete tidal disruption of a planet
\citep{Zubovas2012}; partial tidal disruption of a star
\citep{Guillochon2014} or the corresponding X-ray afterglow emission
originated from the deceleration of the launched jet by the
circumnuclear medium \citep{Yu2011}. These scenarios predict different
long-term activity patterns, which can potentially be inferred from
observations of the reflected signal. Basically, three distinct
patterns might be considered: i) one short\footnote{Here ``short'' and
  ``long'' differentiate between durations shorter or longer than the
  front propagation time over a typical cloud.} flare over the last
$\lesssim 500$ yr, which is associated with a relatively rare event
(and lucky observer), ii) the same but with the longer flare duration of $
\sim 50-100$ yr, iii) a series of short flares separated by few $ 100 $
yr, which can be a manifestation of the secular flaring activity of
Sgr~A* hypothesized from extrapolation of the observed K-band flux
distribution measured over last $ \sim 10$ yrs \citep{Witzel2012} to
rare high-flux events.

Along the same lines, one can extrapolate the distribution of X-ray
flares \citep{2013ApJ...774...42N,2015MNRAS.453..172P} from few Msec
time scales to $\sim$500~yr. To this end we adopted a pure power law
distribution of X-ray fluences $F$ for Sgr~A$^*$ flares:
\begin{eqnarray}
  \frac{dN}{dFdt}\sim 10^{-42}\left(\frac{F}{8\times10^{36}~{\rm erg}} \right )^{-1.5}~{\rm flares\,erg^{-1}\,s^{-1}},
  \label{eq:fluence}
\end{eqnarray}
that characterizes the number of flares $N$ per unit time per unit
interval of fluences\footnote{The observed fluence in ${\rm
    erg\,cm^{-2}}$ has been rescaled to the Sgr~A$^*$ emitted fluence
  from \citet{2013ApJ...774...42N,2015MNRAS.453..172P} assuming 8~kpc
  distance.} In our previous work \citep{churazov16} we estimated the
fluence, required to power the emission from the clouds in the ``Bridge''
region $\displaystyle F \sim 10^{48}\rho_3^{-1}$~erg (here $\rho_3$ is
the mean hydrogen density of the cloud complex in units of $10^3~{\rm
  cm^{-3}}$). From equation~(\ref{eq:fluence}) one can estimate the rate of flares above certain fluence threshold $F$ as $F\frac{dN}{dFdt}$. This estimate predicts approximately one flare with $F\gtrsim 10^{47}\,{\rm erg}$ every 500~yr and one
with $F \gtrsim 10^{48}~{\rm erg}$ every 1500~yr. Of course, the above exercise is very speculative, given that it 
involves an extrapolation over $\sim$10 orders of magnitude in
fluence, which is not directly supported by observations. 

In the short flare scenario, only a small region along the line-of-sight
is illuminated for any given projected position, and it corresponds to
some particular value of the scattering angle cosine $\mu$, so the
polarization degree measurement allows one to determine the radial distance of
the reflecting cloud. The 'nearer vs. further' ambiguity can then be
broken by means of the short-term time variability technique
\citep[][]{churazov16}. The reflected signals from a series of flares can be
distinguished in the same manner, i.e. by decomposing the polarization signal
components with distinct spectral and short-term variability
characteristics. Clearly, this can be done in terms of the $C_1$ and $C_2$
coefficients (see equation~\ref{sec:xipe}) that are linearly related to the
data, and it, in fact, would be an extension of the ``difference''
approach (see \S\ref{sec:dif}) for the polarization data.

For the long flare scenario, projected distance mapping based on
polarization degree and short-term variability becomes inaccurate, therefore some
other technique is needed. Such a technique can be provided by the joint
analysis of the PPV data on
molecular lines with the high-resolution X-ray spectroscopy capable of
measuring the energy (velocity) of the fluorescent lines with
the accuracy of few ${\rm km\,s^{-1}}$. This is possible only with
cryogenic bolometers, like those operated on board ASTRO-H/Hitomi observatory
\citep{2014SPIE.9144E..25T}, and could be done with future missions
like ATHENA and X-Ray Surveyor.  Coupled with polarization and time
variability data \citep[see][]{churazov16}, such analysis would yield a
full 3D density and velocity fields of the molecular gas (and any
weakly ionized material) and would eventually allow for an accurate
reconstruction of the past Sgr~A* X-ray flux over hundreds of year.

\section{Conclusions}
A combination of X-ray imaging, and CCD-type spectroscopy has already
provided us with important information on the distribution of
molecular gas in the Galactic Center region and on the history of the
Sgr~A* activity. Adding polarization information would help to remove
the uncertainty associated with the position of clouds along the line
of sight and eventually to reconstruct the full history of Sgr~A* activity
over few hundred years.

In this paper we show that the variable component in the spectra of
the X-ray bright molecular clouds can be well approximated by a pure reflection
component (see \S\ref{sec:dif}), providing very strong support for the
reflection scenario. We then used a model of the molecular gas 3D
distribution within $\sim 100$~pc from the centre of the Milky Way
\citep{Kruijssen2015} to predict the time evolution and polarization
properties of the reflected X-ray emission, associated with the past
outbursts from Sgr~A$^*$. The model is obviously
too simple to describe the complexity of the true gas distribution in
the GC region. In particular, it does not agree with the approximate
positions of the molecular clouds derived from the variability studies
\citep{churazov16}, although it might still be valid for other
clouds. Future X-ray polarimeters, like \XIPE, have sufficient
sensitivity to tightly constrain the line-of-sight positions of major
molecular complexes (\S\ref{sec:xipe}) and differentiate between
different models. We, in particular, believe that the X-ray brightest
(at present epoch) molecular complex Sgr~A/``Bridge'' is the most
promising target for \XIPE, given its brightness and potentially high
polarization degree. Once the uncertainty in the geometry of the
molecular complexes is removed it should be possible to reconstruct
the long-term variations of the Sgr~A* and place constraints on the
nature of its outbursts.

\section{Acknowledgements}
We thank the referee, Hiroshi Murakami, for useful comments.
The results reported in this article are based in part on data
obtained from the \Chandra X-ray Observatory (NASA) Data Archive and
from the Data Archive of \XMM, an ESA science mission with
instruments and contributions directly funded by ESA Member States and
NASA. We acknowledge partial support by grant No. 14-22-00271 from the
Russian Scientific Foundation. GP acknowledges support by the German
BMWI/DLR (FKZ 50 OR 1408 and FKZ 50 OR 1604) and the Max Planck
Society.

\FloatBarrier

\appendix
\section{CREFL16 model}
\label{ap:model}
\begin{figure}
\includegraphics[trim= 5cm 4cm 0mm 5cm, width=1\textwidth,clip=t,angle=0.,scale=0.49]{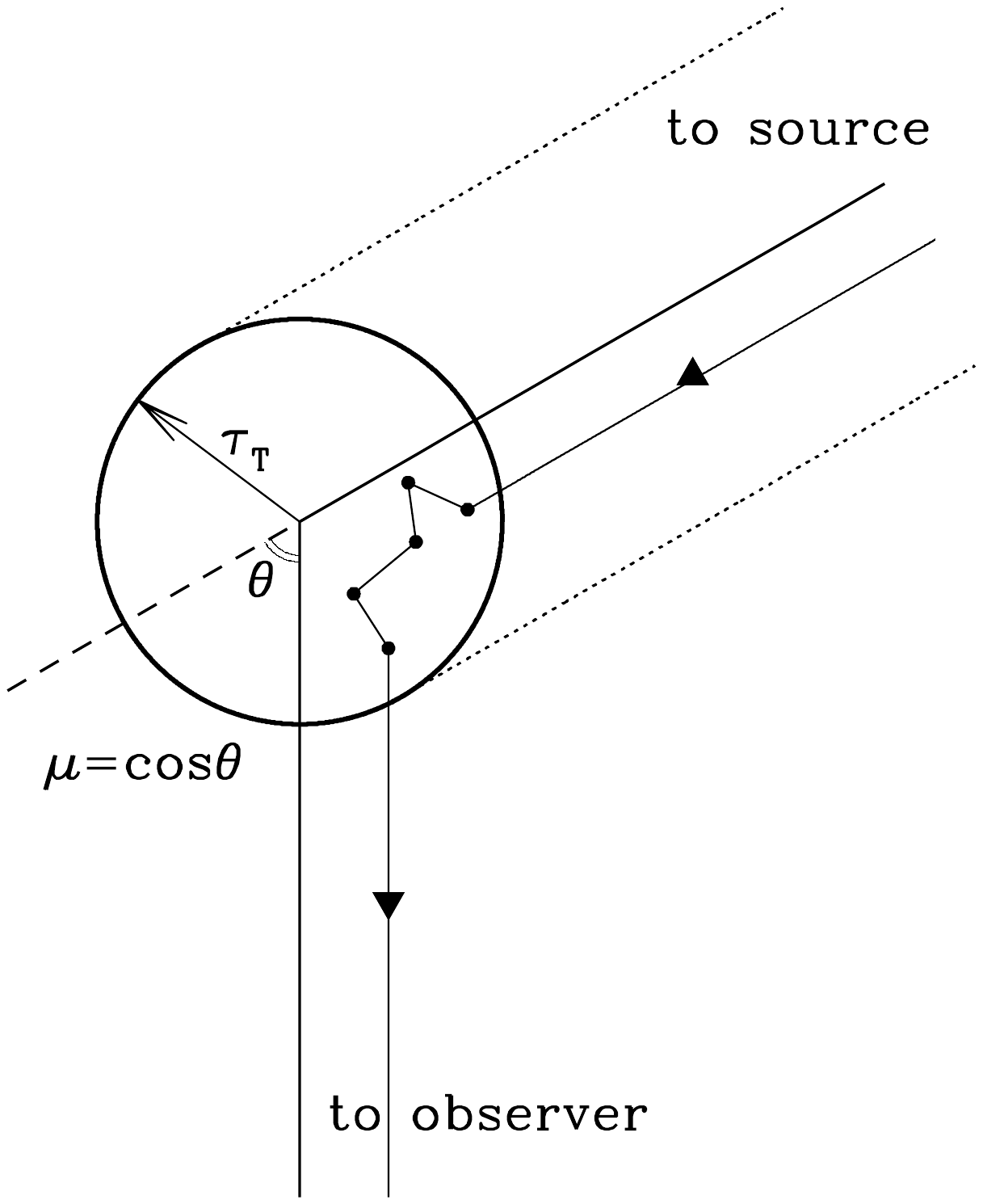}
\caption{Geometry of the problem in the {\small CREFL16} model. A
  spherical homogeneous cloud of cold gas is illuminated by a parallel
  beam of X-rays from a (distant) primary source. The Thomson optical
  depth of the cloud along the radius is $\tau_T$. $\mu$ is the cosine
  of the angle $\theta$ between the line of sight 
  and the direction from the source towards the cloud. The thin solid
  line illustrates a path of a photon experiencing five scattering
  inside the cloud.
\label{fig:geo}
}
\end{figure}

      {\small CREFL16} stands for ``uniform Cloud REFLection model'',
      version of 2016. It is available in a form of the {\small
        XSPEC}-ready table model from
      \verb!http://www.mpa-garching.mpg.de/~churazov/crefl!.  The
      model calculates the reflected spectrum of a uniform spherical
      cloud illuminated by a power law spectrum with a photon index
      $\Gamma$, extending up to 1~MeV {\bf (the geometry of the problem is shown in Fig.~\ref{fig:geo})}. The gas in the clouds is
      neutral and has the abundance of heavy elements (heavier than
      He) set by a parameter $Z$. $Z=1$ corresponds to the abundance
      table of \citet{1992PhyS...46..202F}. The radial Thomson optical
      depth of the cloud is $\tau_T$. The cosine of the viewing angle
      (relative to the direction of primary photons) is set
      by the parameter $\mu$. The range of parameters covered by the
      model is given in Table~\ref{tab:par}. The energy range covered
      by the model is 0.3-100 keV, with logarithmic grid over energy
      (584 points).
      
\begin{table}
  \caption{Range of parameters covered by the {\small CREFL16} model.
    \label{tab:par}
  }
  \begin{tabular}{c | c | c}
    \hline
    Parameter & min & max \\
    \hline
        $\tau_T$ & $10^{-4}$  & 13 \\
        $\Gamma$ & 1.2  & 2.6 \\
        $Z$ & 0  & 3.0 \\
        $\mu$ & -1  & 1 \\
    \hline
  \end{tabular}
\end{table}

\label{lastpage}
\end{document}